\documentclass{article}
\PassOptionsToPackage{table}{xcolor}
\usepackage[utf8]{inputenc}
\usepackage[T1]{fontenc}
\usepackage[a4paper, total={5.5in, 8in}]{geometry}
\usepackage{url}
\usepackage{booktabs}
\usepackage{amsmath, amssymb, amsfonts}
\usepackage[numbers,super,sort&compress]{natbib}
\renewcommand{\thefootnote}{\alph{footnote}}
\usepackage{array}
\usepackage{nicefrac}
\usepackage{microtype}
\usepackage{float}
\usepackage{caption}
\usepackage{lscape}
\usepackage{setspace}
\usepackage{eurosym}
\usepackage{dcolumn}
\usepackage{multirow}
\usepackage{subcaption}
\usepackage{ragged2e}
\usepackage{todonotes}
\usepackage{threeparttable}
\usepackage{longtable}
\usepackage{graphicx}
\usepackage{xcolor}
\usepackage{colortbl}
\definecolor{pos}{RGB}{220,245,220}
\definecolor{neg}{RGB}{245,220,220}
\definecolor{mix}{RGB}{245,240,220}
\definecolor{nul}{RGB}{235,235,235}
\usepackage[colorlinks=true,
            linkcolor=blue,
            urlcolor=blue,
            anchorcolor=blue,
            citecolor=blue]{hyperref}
\providecommand{\keywords}[1]{%
  \small
  \textbf{\textit{Keywords---}} #1
}

\makeatletter
\newcommand{\titlethanks}[1]{%
  \begingroup
    \renewcommand\thefootnote{*}%
    \footnotetext[0]{#1}%
  \endgroup
}
\newcommand{\affilthanks}[1]{%
  \begingroup
    \renewcommand\thefootnote{ }%
    \footnotetext[0]{#1}%
  \endgroup
}
\makeatother

\title{Sources of Inequality at Birth: The Interplay Between Genes and Parental Socioeconomic Status\textsuperscript{*}\titlethanks{Research reported in this publication was supported by NORFACE DIAL funding (462-16-100), the National Institute On Aging of the National Institutes of Health (RF1055654, R56AG058726, R01AG078522, and R01AG079554), the Dutch National Science Foundation (016.VIDI.185.044), and the European Social Science Genetics Network (ESSGN) under the European Union's Horizon 2020 research and innovation programme's Marie Sk\l{}odowska-Curie grant agreement (ESSGN 101073237). The views expressed in this publication are those of the authors and do not necessarily reflect those of the European Union, MSCA Horizon Europe, or ESSGN. Neither the European Union, nor the granting authority or ESSGN can be held responsible for them.}%
\affilthanks{\textsuperscript{1}University of Bologna;\quad
\textsuperscript{2}Vrije Universiteit Amsterdam;\quad
\textsuperscript{3}Tinbergen Institute;\quad
\textsuperscript{4}Erasmus University Rotterdam;\quad
\textsuperscript{5}University of Zurich;\quad
\textsuperscript{6}University of Iowa;\quad
\textsuperscript{7}University of Bristol;\quad
\textsuperscript{8}Northwestern University;\quad
\textsuperscript{9}University of Southern California.}}

\author{%
Pietro Biroli\textsuperscript{1,\,\S} \and
Nicolau Martin-Bassols\textsuperscript{1,\,\S} \and
Andries T. Marees\textsuperscript{2} \and
Hans van Kippersluis\textsuperscript{3,4} \and
Cornelius A. Rietveld\textsuperscript{3,4} \and
Pia Arce\textsuperscript{5} \and
Kevin Thom\textsuperscript{6} \and
Stephanie von Hinke\textsuperscript{7} \and
Jeremy Vollen\textsuperscript{8} \and
Titus Galama\textsuperscript{2,3,9,\,$\dagger$}
\\[1ex]
\normalfont\footnotesize $\S$ Shared first author. \quad $\dagger$ Last author, Corresponding author.%
}
\date{}

\begin{document}

\maketitle

\begin{abstract}
The start of a human's life can be characterized by two lotteries: that of your genes (nature) and the family you were born into (nurture). These set in motion a trajectory, from birth onward, in health and human capital. Leveraging three longitudinal social-science data sets, we systematically analyze the relationship between an individual's genotype, the socioeconomic status (SES) of the families they grew up in, and their realized traits in adulthood. We proxy an individual's genetic predisposition by polygenic indexes (PGIs) and family SES by a latent factor of parental education and father's (former) occupational status.  We then investigate how PGIs, parental SES, and their interaction contribute to later-life outcomes across a range of forty-five socioeconomic, anthropometric, health, behavioral, and personality traits.  We find strong genetic and socioeconomic associations with these phenotypes, but no evidence of sizable gene-environment interactions.
\end{abstract}
 \\

\newpage 

\doublespacing

\section{Introduction}

Families play an essential role in shaping life chances, as evidenced by the strong intergenerational transmission of advantage \citep{BlauDuncan1967,Jencks1972, BourdieuPasseron1977, Becker1986,Black2005,Mare2011, Solon1992, Shonkoff2000, Heckman2006,Corak2013,Chetty2020}. 
Indeed, the start of a human’s life can be characterized by two lotteries: one assigning genes (nature), and the other assigning a rearing environment (nurture) \citep{Harden2021book,houmark2024nurture}. Both operate through familial ties, and each has a substantial influence on life chances. 
Twin studies suggest that for a wide variety of traits, some 30 to 70 percent of the variation between individuals can be attributed to differences in genetics \citep{polderman2015meta}. However, heritability estimates in this range imply that while genes matter, environments also play an important role.  One particularly important environmental factor is parental socioeconomic status (SES), the social position that parents occupy based on education, occupation, income, and related characteristics.  Parental SES is strongly tied to a family's economic resources, and their ability to invest in the development and well-being of their children.  A large literature studies how familial SES shapes early-life conditions and the subsequent adult outcomes of children  \citep[][]{barker1993fetal,galobardes2008association,cunha2006interpreting,Almond2011vol1,Aizer2014intergen,Almond2018vol2}.

A central question is whether the two ``lotteries'' over genes and family SES interact in shaping realized outcomes. 
The literature on gene-environment (G$\times$E) interactions explores whether certain environments reduce or exacerbate genetic advantage \citep{Haldane1946,plomin1977genotype,mcallister2017current, biroli2026economics}.  There are strong theoretical motivations to investigate this in the specific context of parental SES.  In economics, child endowments (including genes) and parental resources are often considered distinct inputs in the production of children's outcomes \citep{Becker1986}.  Yet, it is not clear whether they act as substitutes or complements. That is, parental resources could compensate for less advantageous endowments, or they could multiply the benefits of advantageous endowments.  In the behavioral genetics literature, a prominent theory is the Scarr--Rowe hypothesis \citep{scarr1971race,rowe1999genetic}, which posits that genetic influences on cognitive ability are more fully expressed in advantageous environments.  Empirically evaluating these hypotheses is important.  Many redistributive social policies attempt to mitigate disparities in familial SES.  The presence of gene-SES interactions suggest that such policies may also moderate gradients induced by the genetic lottery. 

No firm consensus has emerged from the empirical literature evaluating gene-SES interactions.  Empirical tests using twin and sibling designs have produced heterogeneous findings across cohorts and national contexts \citep{rowe1999genetic, Turkheimer2003, harden2007genotype, Hanscombe2012, tucker2016large}. 
In some settings, genetic variance appears to be lower under socioeconomic disadvantage. In others, such moderation is weak or absent. These results have been interpreted as reflecting complementarity between genetic endowments and environmental resources in some contexts, and substitutability in others.

One path forward in this literature involves the use of observed measures of genetic factors constructed from molecular genetic data.  The growing availability of molecular genetic data, and the development of new statistical methods for exploring G$\times$E interactions allow researchers to explore these questions in larger and richer social-science datasets \citep{Domingue_2020B,biroli2026economics}.   Genome-wide association studies (GWASs) have robustly identified relationships between specific single-nucleotide polymorphisms (SNPs; a common measure of genetic variation) and a large and growing number of health and human-capital related outcomes \citep{hill2016molecular, lee2018gene, yengo2018meta, abdellaoui2023}. These molecular-level associations can be aggregated into Polygenic Indices (PGIs) that can account for a substantial proportion of the variation in their targeted phenotypes \citep{bulik2015atlas,becker2021resource}.  

A growing literature explores interactions between PGIs and a variety of environmental exposures, including adverse conditions during childhood, neighborhood characteristics, and policies like cigarette taxes or compulsory schooling laws \citep{Meyers2013,barcellos2018education,Domingue2015, Fletcher2012,Schmitz2016,Treur2017,bierut2018childhood}.  While many of these results provide evidence specifically on gene-SES interactions, the robustness of the evidence varies depending on the traits studied, and the aspect of SES that is being measured.   Although Bierut et al. (2023)\cite{bierut2018childhood} find evidence that childhood financial adversity exacerbates genetic risk for smoking, the review by Pasman et al. (2019)\cite{pasman2019systematic} could not draw firm conclusions about genetic interactions with related environments (e.g., childhood trauma and parental education) for a wider range of substance use.  Similarly, for psychiatric traits, such as major depressive disorder, interactions between PGIs and environmental variables such as childhood trauma are modest and sometimes produce conflicting results (see Assary et al. 2018\cite{assary2018gene} for a review). More robust findings exist for physical health outcomes, with social deprivation exacerbating the genetic risk of obesity \citep{Tyrrell2017, barcellos2018education, amin2017gene, hoang2023heterogenous}.  In the case of educational attainment (EA), there is some evidence that the relationship between an EA PGI and college completion is stronger among higher SES families \citep{papageorge2020genes,ghirardi2024Interaction, ghirardi2025Compensating}.

There thus exists some evidence that G$\times$SES interactions may shape health- and human capital-related outcomes. However, the fragmented nature of the evidence leaves open the question of whether these interactions constitute a general feature of genetic associations or are instead specific to particular traits and environments. This uncertainty is driven by three features of the literature.  First, the strength of the evidence varies markedly across domains and measures of SES, with more robust findings for some behavioral and physical health outcomes than for psychiatric disorders \citep{Domingue2015, Fletcher2012, Meyers2013, Schmitz2016, Treur2017, bierut2018childhood,assary2018gene}. This cross-domain heterogeneity suggests that G$\times$E associations may be highly trait-specific. Second, the range of outcomes analyzed remains limited, as most studies focus on a small set of traits without systematically assessing whether similar interaction patterns emerge elsewhere. Results may therefore also reflect publication bias. 
Third, existing studies are typically conducted within single datasets, making it difficult to determine whether reported interactions are stable across samples or instead reflect particular environments. 


Our study seeks to address these challenges by systematically estimating and comparing the magnitude of the interaction between a consistently measured measure of childhood environment (parental SES) -- proxied by a latent factor of years of education of the father, years of education of the mother, and the occupational status of the father -- and a set of consistently measured genetic predispositions (measured by PGIs) on the outcome of interest measured in adulthood, for 45 traits related to health and human capital, in three longitudinal social science data sets: the Health and Retirement Study (HRS), the Wisconsin Longitudinal Study (WLS), and the English Longitudinal Study of Ageing (ELSA) \citep{Sonnega2014cohort,herd2014cohort,Steptoe2013}. 


\section{Results}

Using a mega-analysis that pools data from three datasets (HRS, WLS, and ELSA), we explore how genetic predispositions $PGI$, socioeconomic background $SES$, and their interaction $PGI \times SES$ contribute to later-life outcomes running 45 OLS regressions, one for each trait, in each of the three datasets, as:
\begin{eqnarray}\label{eq:OLS_gxe}
{Y^{s}_{i} =  \alpha^s + \beta^s PGI^{s}_i +  \gamma^s SES_{i}^{'} + \rho^s \left[PGI^{s}_{i} \times SES_i \right] + \mathbf{X}_{i}^{'} \delta_s + \varepsilon_{i}^{s}, \nonumber}
\end{eqnarray}
where $Y^{s}_{i}$ are the different socioeconomic, anthropometric, health, behavioral, and personality-related traits indexed by $s$ (where $s$ ranges over 45 different outcomes); 
$PGI^{s}_i$ are the outcome specific polygenic indices (each PGI is constructed to predict its phenotype); 
and $SES_{i}$ (parental SES) is a proxy for the SES of the family one was born into. 
We control for $X_{i}$, a vector containing an indicator for the dataset (HRS, WLS, and ELSA), the first ten ancestry-specific principal components of the genetic data, age and age squared (in years), 
sex, region of birth, interactions between age  and sex, and interactions of all baseline controls with the PGI and parental SES.

\begin{figure}
    \centering
    \includegraphics[scale=0.45]{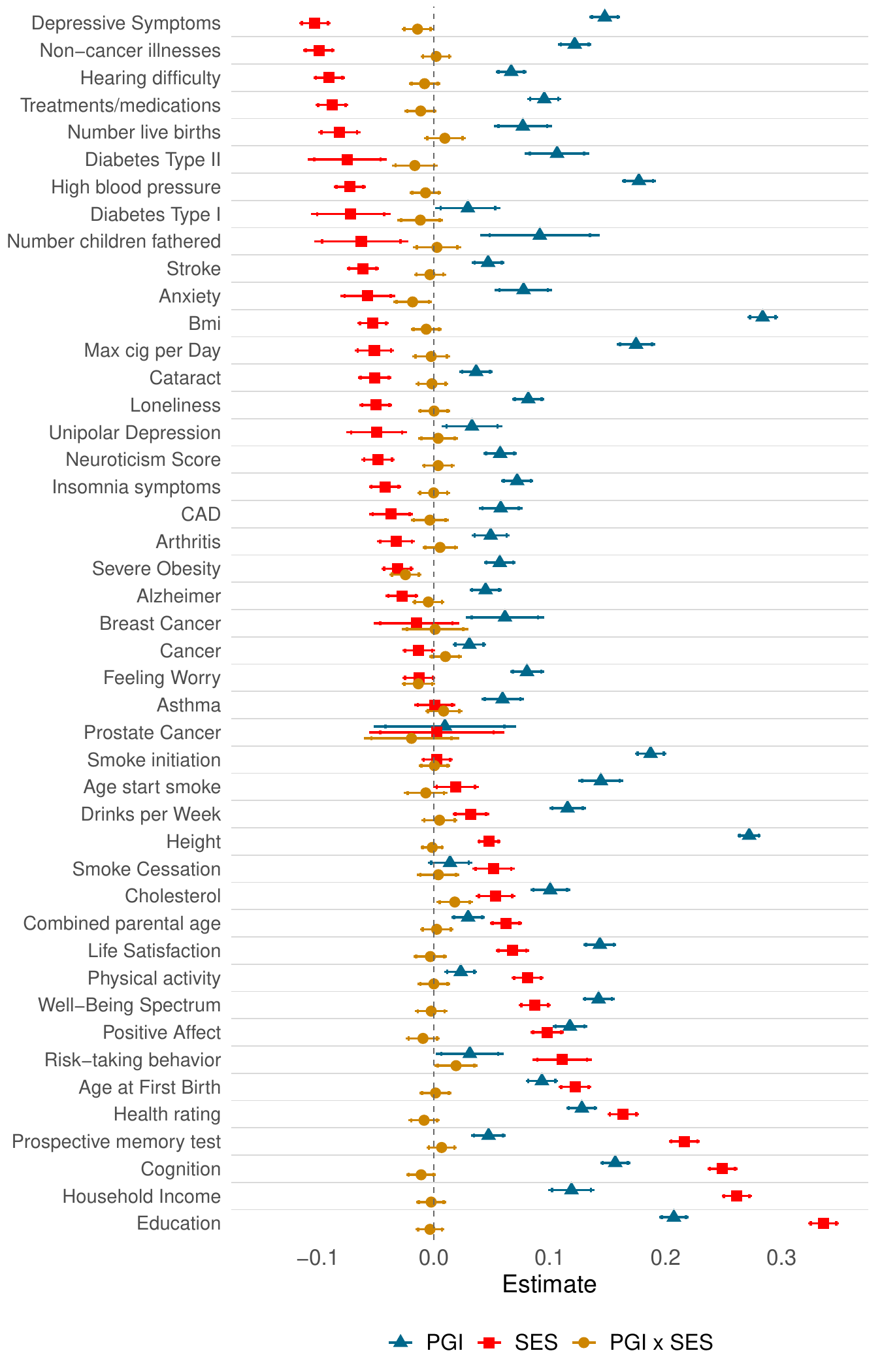}
\caption{Mega-analysis: Associations between PGIs, parental SES, and PGI × SES across phenotypes}
    				 \justifying 
	\vspace{0.2cm}
    \noindent \footnotesize{Plot of the coefficients estimated from the OLS regression model specified in \autoref{eq:OLS_gxe}, showing the associations along with their 95\% confidence intervals, between polygenic indices (PGIs; triangles), parental socioeconomic status (SES; squares), and their interaction (PGI $\times$ SES; circles) across 45 phenotypes. The mega-analysis pools all three datasets, the HRS, the WLS, and ELSA. Control variables include an indicator for each dataset, 10 ancestry-specific principal components of the genetic data, age and age squared, sex, region of birth, and interactions between age and sex.}
    \label{fig:ols_gxses_pgs1_all}
\end{figure}

The main coefficients of interest are $\beta^s$, capturing the extent to which the PGIs are predictive of their respective outcomes, $\gamma^s$, denoting the relation between parental SES and the outcomes, and $\rho^s$, estimating the interaction between the PGI and parental SES.
The estimated coefficients from the mega-analysis pooling all three datasets are plotted in Figure \ref{fig:ols_gxses_pgs1_all} for all 45 traits.
We find strong genetic and socioeconomic associations: both $\beta^s$ and $\gamma^s$ are large in magnitude and statistically different from zero for most outcomes. 
Specifically, genetic associations $\beta^s$ (blue triangles) are positive and significant (except for Type I diabetes and smoking cessation, which are close to zero). SES associations $\gamma^s$ (red squares) exhibit a different pattern, with qualitatively adverse conditions such as depressive symptoms and anxiety consistently presenting negative associations with the SES measure, and qualitatively advantageous outcomes such as educational attainment and household income presenting positive associations, showing that parental SES is associated with better health and higher SES in later life. These findings underscore the effective predictive power of both PGIs and the parental SES measure. In contrast, we find no evidence of economically meaningful interactions between the two (orange circles): the estimated $\rho^s$ coefficients are small in magnitude and centered around zero.\footnote{The full set of PGI, SES, and PGI $\times$ SES estimates with standard errors for all 45 phenotypes is reported in Supplementary Table \ref{tab:coef45}.  Corresponding dataset-specific estimates for the HRS, the WLS, and ELSA are reported in Appendix Figure \ref{fig:ols_gxses_pgs1_all1}. Results across the datasets are highly consistent.}

\begin{figure}
    \centering
    \includegraphics[scale= 0.60]{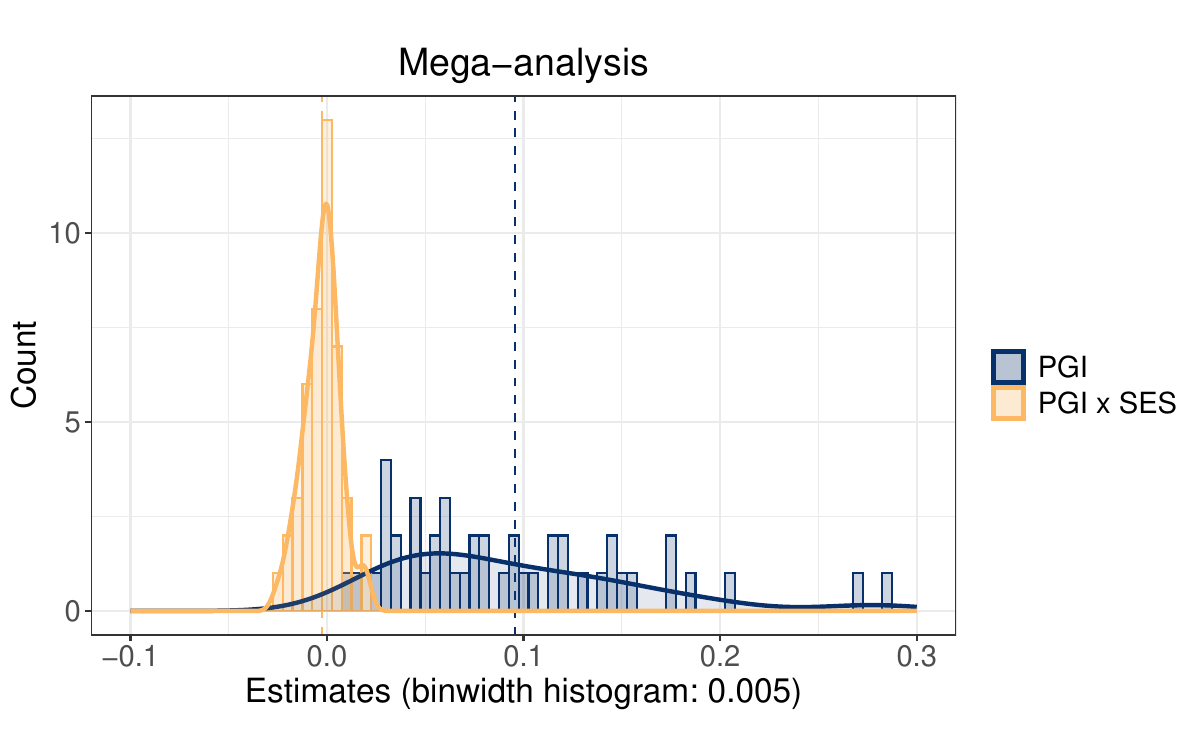}
        \caption{Density and histogram of PGI and PGI × SES coefficients across 45 phenotypes}
    				 \justifying 
    \vspace{0.2cm}
    \noindent\footnotesize{Smoothed density and overlaid histogram representing the distribution of estimated coefficients from the OLS regression model specified in \autoref{eq:OLS_gxe}, based on 45 estimated coefficients, one for each phenotype. These coefficients are pooled from the Health and Retirement Study (HRS), the Wisconsin Longitudinal Study (WLS), and the English Longitudinal Study of Ageing (ELSA). The dashed vertical line represents the average coefficient for PGI and PGI $\times$ SES across the different phenotypes.}
    \label{fig:density_histo_pooled}
\end{figure}

To better illustrate this, Figure \ref{fig:density_histo_pooled} plots the distribution of the PGI, $\beta^s$, and interaction PGI$\times$SES, $\rho^s$, coefficients across the 45 different outcomes, pooled across the three datasets. The PGI coefficients are always positive, with an estimated magnitude of around 10\% of a standard deviation ($\beta^s=0.1$), while the coefficients of the PGI$\times$SES interactions are always close to zero, $\rho^s \approx 0$. 
This suggests that the statistical interaction between genes and parental SES is not  substantial.

\begin{figure}
    \centering
    \includegraphics[scale= 0.6]{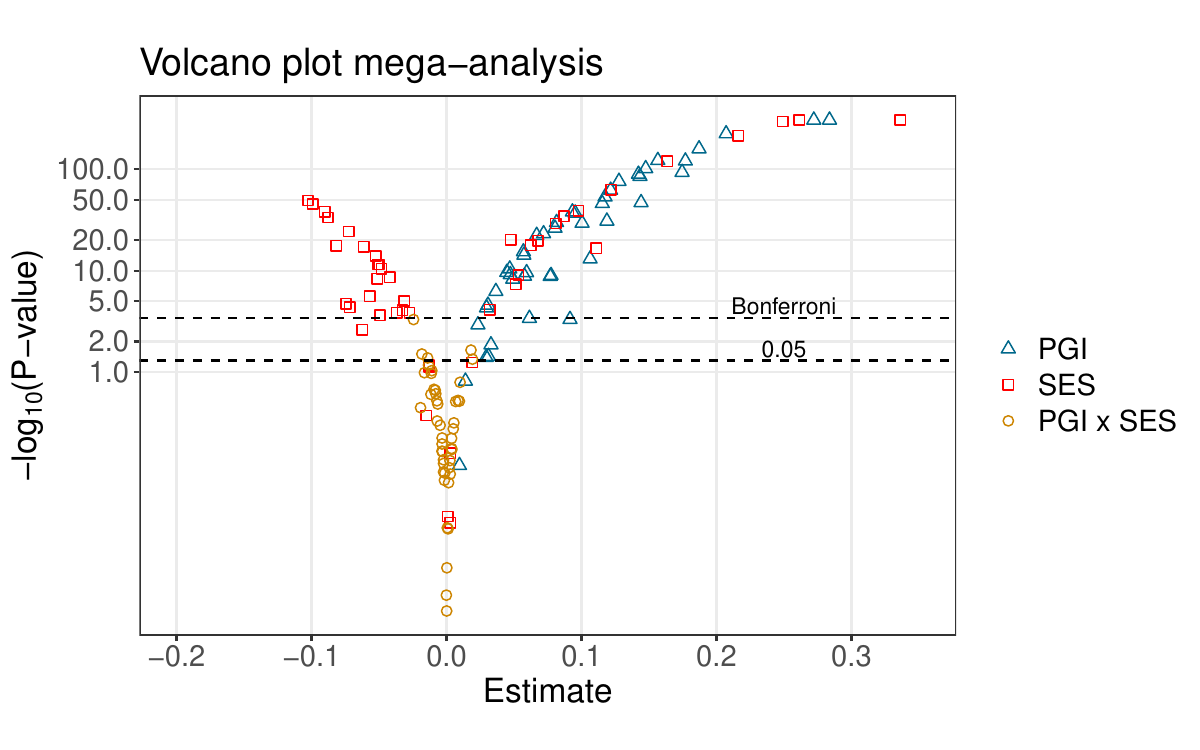}
    \caption{Volcano plot of PGI, parental SES, and PGI $\times$ SES coefficients across 45 phenotypes}
    \justifying 
    \vspace{0.2cm}
    \noindent
    \footnotesize{Estimated coefficients from the OLS regression model specified in \autoref{eq:OLS_gxe} plotted against their corresponding -log10(p-values), pooling the HRS, the WLS, and ELSA. The horizontal reference lines indicate conventional significance thresholds and Bonferroni-adjusted cutoffs for multiple hypothesis testing.}
    \label{fig:volcano}
\end{figure}

Moreover, as shown in Figure \ref{fig:volcano}, the coefficients $\rho^s$ for the PGI$\times$SES interactions (orange circles) are virtually all statistically indistinguishable from zero. Only five interaction terms are significantly different from zero at the 95\% confidence level, namely those for risk-taking behavior, cholesterol, severe obesity, anxiety, and depressive symptoms. None of these findings remain significant after Bonferroni adjustment for multiple hypothesis testing. 
In contrast, both the PGI (blue triangles) and parental SES (red squares) coefficients remain statistically significant after Bonferroni correction, with exceptions in only six (parental SES) and eight (PGI) instances, respectively.

\begin{figure}
\centering
    \includegraphics[scale= 0.6]{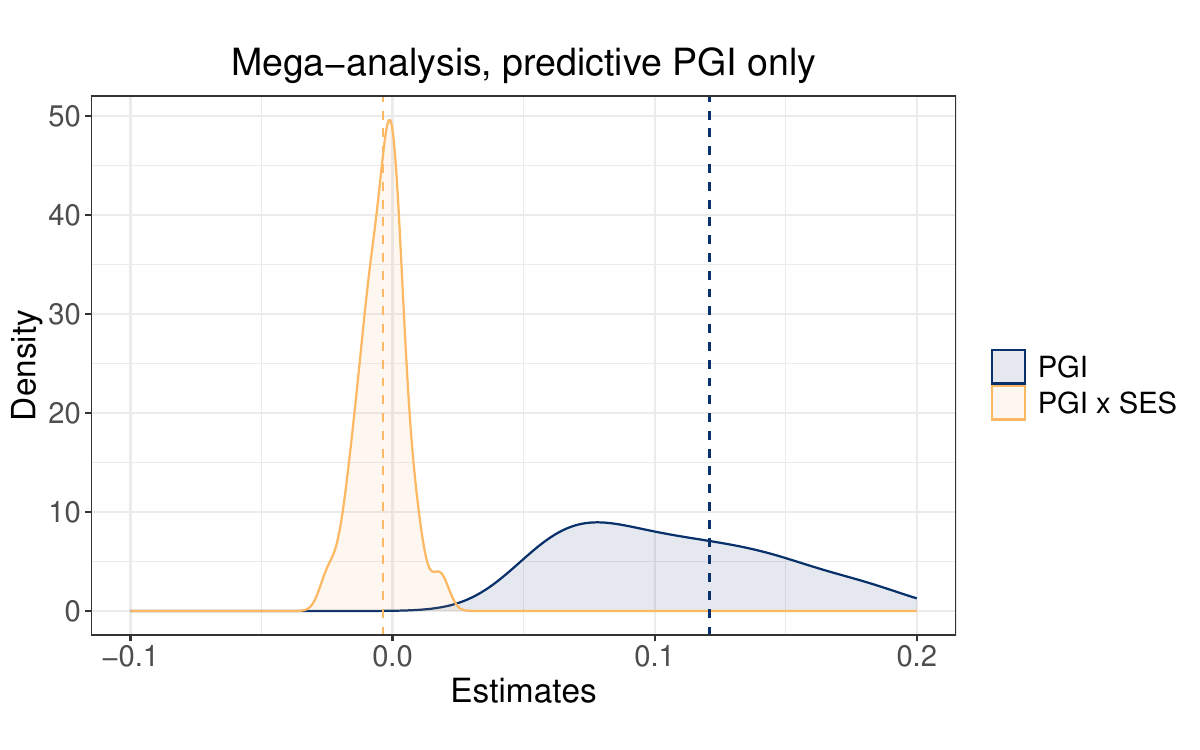}
        \caption{Histogram of estimated coefficients for PGI and PGI $\times$ SES}
        \justifying 
    \vspace{0.2cm}
    \noindent
    \footnotesize{Smoothed density plot of the estimated coefficients for PGIs and their interaction with SES (PGI $\times$ SES). Pooled OLS regression model specified in \autoref{eq:OLS_gxe}, limited to PGIs with greater predictive power, where coefficients ($\beta$) are greater than 0.05.} 
        \label{fig:density_005}
\end{figure}

To rule out the concern that the estimated interaction terms $\rho^s$ are close to zero due to limited predictive power of PGIs, we next select traits with relatively large genetic associations ($\beta^s > 0.05$). Figure \ref{fig:density_005} presents the distribution of the corresponding PGI ($\beta^s$) and PGI$\times$SES interaction ($\rho^s$) coefficients. The interaction estimates remain small and tightly clustered around zero, indicating the near-zero interactions are not due to weak PGIs. Even for the PGIs with the strongest predictive power, such as educational attainment, the interaction remains close to zero. For education, the estimated PGI main effect is $\beta^s = 0.207$ ($p < 0.001$), the parental SES main effect is $\gamma^s = 0.336$ ($p < 0.001$), and the PGI $\times$ SES interaction is $\rho^s = -0.003$ ($p = 0.598$) --- small and statistically indistinguishable from zero. Imposing an even more stringent criterion of restricting the analysis to phenotypes with large main associations for both the genetic predisposition and for parental SES ($\beta^s > 0.10$ and $\gamma^s > 0.10$), we end up with only six traits (13.3\%) that meet this threshold. Yet even within this subset, the estimated PGI$\times$SES interactions remain statistically indistinguishable from zero.

\begin{figure}
    \centering
    \includegraphics[scale= 0.5]{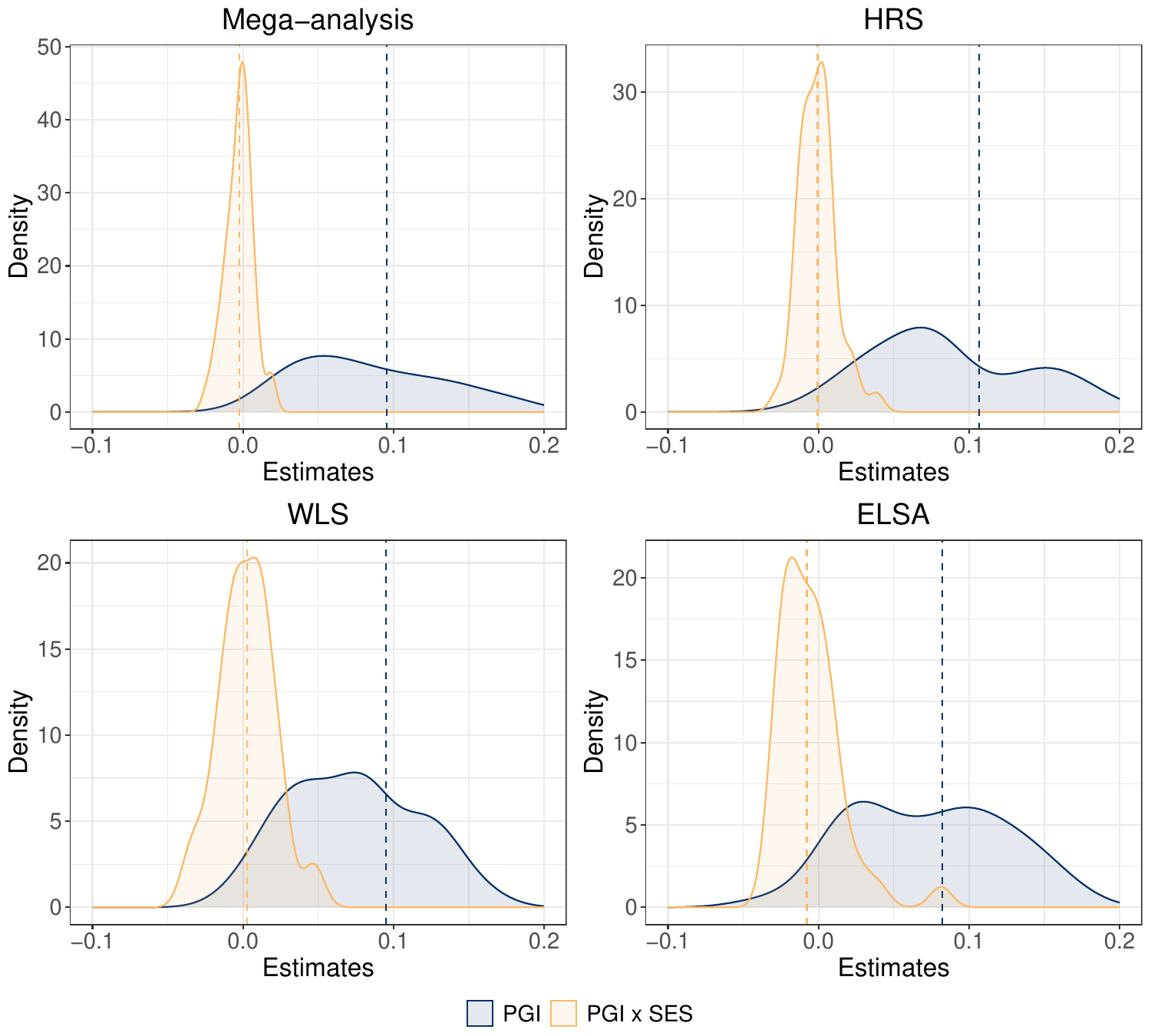}
         \caption{Histogram of PGI and PGI $\times$ SES coefficients across 45 phenotypes for different databases}
    				 \justifying 
    \vspace{0.2cm}
    \noindent \footnotesize{Smoothed density of the estimated coefficients for the PGI and PGI $\times$ SES from the OLS regression model specified in \autoref{eq:OLS_gxe}, presented separately for each dataset: the HRS, the WLS, and ELSA. The density plots illustrate the distribution of estimated coefficients within each dataset, allowing for a comparison of their spread and central tendency across the three samples. The dashed vertical line represents the average coefficient for PGI and PGI $\times$ SES.}
    \label{fig:density_all}
\end{figure}

All three datasets sample individuals in middle- and older-age, but while the HRS is nationally representative of the US population older than 50, the WLS is representative of Wisconsin’s high school class of 1957 and of similarly aged white Americans who completed high school, and the ELSA is representative of the English population older than 50. These differences in sampling and contexts could lead to heterogeneity in the interaction terms, potentially averaging out to near-zero associations. Figure \ref{fig:density_all} however, suggests this is unlikely. Across all datasets, PGIs exhibit similar predictive power, centered around $\beta=0.1$. Importantly, the estimated interaction terms are consistently centered around zero, albeit with greater confidence intervals due to smaller sample size, indicating that the absence of substantial interactions is not specific to any particular sample.

\section{Discussion}

Despite an extensive literature showing evidence of G$\times$E interactions, the question of their existence, for which traits, genotypes and environments, remains far from resolved, with numerous ambiguous and even conflicting results, e.g.,\cite{Domingue2015, Fletcher2012, Meyers2013, Schmitz2016, Treur2017, bierut2018childhood, pasman2019systematic, assary2018gene, Tyrrell2017, barcellos2018education,amin2017gene, hoang2023heterogenous, papageorge2020genes}. We approached this question by systematically exploring PGI$\times$SES interactions in three longitudinal social-science data sets containing information on 45 human capital-related phenotypes, 45 related genotypes, and parental SES. Utilizing these data, we assessed the predictive power of parental SES, of PGIs, and of the linear interaction between parental SES and PGIs across the 45 phenotypes. At a 95\% confidence level, PGIs significantly predict 43 of 45 phenotypes (95.6\%) and parental SES significantly predicts 38 of 45 phenotypes (84.4\%), whereas statistically significant PGI$\times$SES interaction associations are observed for only 5 of 45 phenotypes (11.1\%). None of these interaction terms remain statistically significant after correcting for multiple hypothesis testing. Importantly, this finding is not driven by weak genetic or weak SES gradients. Restricting the analysis to phenotypes with statistically significant and economically meaningful main associations, we still find no statistically significant PGI$\times$SES interactions. Taken together, these findings provide little evidence of meaningful linear G$\times$E interaction between PGIs and parental SES.

In other words, the predictive power of the PGIs we use does not appear to systematically differ between individuals born into high and low SES families. Thus, using a DNA-based measure of genetic predispositions, we do not find support for the Scarr-Rowe hypothesis. Previous PGI$\times$E interaction studies have shown conflicting results regarding the Scarr-Rowe hypothesis. US-based studies confirmed the hypothesis \citep{papageorge2020genes}, whereas European-based studies found the interactions to be nonexistent or even reversed \citep{isungset2022social}. This difference between US and European settings is in line with twin research that showed that Scarr–Rowe interactions are non-existing in countries with broader welfare policies \citep{tucker2016large}. However, in the present study, exploiting both US and UK data, we find negligible differences across datasets.

There are two main interpretations of our work. First, the lack of significant PGI$\times$SES interactions in a systematic and comprehensive analysis could indicate that previous results were affected by `publication bias', the tendency for studies with significant results to be published more frequently than studies with insignificant results \citep{sterling1959publication}. This bias can skew the overall literature on a topic, as the published research may not accurately represent the true effect size or the range of results. This has been found to be a prevalent issue in the social sciences \citep{franco2014publication, peplow2014social}. In the young field of social genomics, this bias may have led scholars, consciously or unconsciously, to concentrate on a select group of phenotypes and genetic measures that present G$\times$E interactions, thereby neglecting a broader spectrum of relevant lifetime outcomes. In an aggregate manner, this phenomenon could have contributed to the impression that G$\times$E interactions are the norm, while in reality, they may be the exception. This misperception could skew both theoretical understanding and practical applications, leading to an overestimation of the prevalence and significance of G$\times$E interactions in the literature. This interpretation, consistent with our results, would suggest that overall, genes (nature) and parental SES (nurture) contribute to human capital formation \citep{turkheimer2000three, polderman2015meta,hout1993persistence, siponen2011children}, but mostly in an independent manner.

However, despite using some of the largest GWASs available, it could be that our genetic measures simply have too much measurement error \citep{van2023overcoming}.
Moreover, the parental SES measure used in the current study might not be optimal: it is possible that the size of a  PGI$\times$SES interaction is not related to general parental SES, but to rather more extreme or specific situations such as severe financial distress or an abusive family environment. Studies have shown that childhood maltreatment and traumatic circumstances can have large effects on children, lasting into adulthood \citep{norman2012long,gardner2019association,d2022childhood}. Epigenetic studies even show clear biological mechanisms explaining life course differences between children with and without a history of maltreatment \citep{teicher2016effects,cowell2015childhood,teicher2016annual}.

Not finding systematic PGI$\times$SES may also be the result of incorrectly specifying the underlying model: human capital may be a more complex function of both nature and nurture, not captured by a simple model with a linear interaction term. Therefore, the PGI may fail to capture the specific measures of genetic variation underlying the interplay between genetic variation and SES. Potentially, methods such as the variance GWAS \citep{johnson2022polygenic,schmitz2021impact} or PIGEON \citep{miao2025pigeon} may elucidate more complex interaction mechanisms, as such approaches utilize more direct measures of variance in phenotypic outcomes associated with SNPs. Finally, parental SES and genetic inheritance are not independent in a between-family setting. This gene-environment correlation ($rGE$) may bias the estimation of PGI$\times$SES interactions \citep{biroli2026economics}. 

Another possibility is that G×E estimates are sensitive to outcome definition, as alternative specifications group individuals into substantively distinct categories that may be differentially affected by both SES and genetic factors. This is illustrated by the interaction with educational attainment, where our results diverge from those of Papageorge and Thom (2020)\cite{papageorge2020genes} and  Ghirardi and Bernardi (2025)\cite{ghirardi2025Compensating}, despite relying on the same data (HRS), phenotype (education), and genetic measure (EA PGI). The discrepancy appears to arise from differences in how the outcome is constructed: our analysis uses years of education as a continuous measure, whereas Papageorge and Thom (2020)\cite{papageorge2020genes} and  Ghirardi and Bernardi (2025)\cite{ghirardi2025Compensating} focus on dichotomous outcomes (e.g., college or high school completion). Replicating their approach, we find a positive and statistically significant interaction between the education PGI and SES when using a college completion indicator ($\rho = 0.01$, $p = 0.007$). In contrast, when high school completion is used as the outcome, the interaction is negative and statistically significant ($\rho = -0.06$, $p = 0.000$). This pattern suggests that college completion is a highly selective margin along which SES amplifies genetic effects, whereas high school completion reflects a less selective threshold where SES attenuates them. Consequently, aggregating these distinct margins into a linear measure such as years of education may mask offsetting effects and produce null estimates. More broadly, these results highlight the sensitivity of G×E estimates to outcome definition and underscore the importance of adopting outcome-specific, theoretically grounded measures for which interactions are substantively meaningful.

These potential interpretations are not mutually exclusive and likely explain different aspects of our results. Yet, notwithstanding the methodological limitations, the evidence presented here meaningfully shifts prior expectations by suggesting that G$\times$E interactions may be less widespread across major life-course outcomes than previously assumed, with genetic and socio-economic factors more often acting independently. Our findings also underscore the necessity of advancing genetic measures with enhanced predictive capabilities and identifying specific environmental factors influencing genetic expression. Furthermore, it emphasizes the importance of developing models and functional forms that best elucidate the interactions between genes and the environment.

\pagebreak
\section{Methods}
\subsection{Data}
We use three longitudinal social science datasets,  the Health and Retirement Study (HRS), the Wisconsin Longitudinal Study (WLS), and the English Longitudinal Study of Ageing (ELSA), containing rich genotypic and phenotypic information. We systematically analyze the relationship between the offspring's (henceforth, the child's) genetic propensity towards a trait (genotype), the socioeconomic status (SES) of the families they grew up in when they were children, and their actual trait in adulthood (phenotype).

The HRS is a longitudinal household survey providing rich data on about 26,000 individuals, representative of the U.S. population over the age of 50. Up to 13 waves of data per respondent are available, from 1992--2016. We use the publicly available HRS core survey and linked genetic data for the years 2006, 2008, and 2010. HRS core surveys are conducted biennially using a combination of face-to-face and telephone interviewing. 

The WLS is a long-term study of a sample of 10,317 men and women who graduated from Wisconsin high schools in 1957. Survey data were collected from the original respondents or their parents in 1957, 1964, 1975, 1992, 2004, and 2011.

ELSA is a study on the dynamics of health, social, well-being, and economic circumstances in the English population aged 50 and older. More than 18,000 people have taken part in the study since it started in 2002, with the same people re-interviewed every two years.

\subsection{Phenotypes}

Phenotypes were selected from the GWAS catalog and the GWAS Atlas \citep{GWASCatalog_2013,watanabe2019global}, versions May-30-2019 and April-17-2019, respectively. We selected the most relevant and powerful GWASs for various trait `domains/families'. 
All included GWASs passed the following exclusion criteria:
\begin{itemize}
    \item Number of SNPs > 450,000.
    \item Over 80 percent of samples of EUR population.
    \item SNP-based heritability Z-score > 2.
    \item All relevant data fields available (e.g., beta or odds ratio, effect, and non-effect alleles).
    \item GWAS with the largest $N$ for the trait of interest.
    \item The exact or highly comparable phenotype must be available in the HRS, WLS, and ELSA.
\end{itemize} 

\vspace{5mm} 

Ultimately, we selected 45 traits (phenotypes), with corresponding GWASs. An overview of these 45 GWASs is provided in Supplementary Table \ref{tab:gwas_references}.

We defined each trait across HRS, WLS and ELSA, constructing 45 phenotypes that are as similar as possible to the definition of the corresponding GWAS. In cases where the  phenotypes differed from the GWAS definition, we harmonized the phenotypes as much as possible.       
Phenotypes between studies were sometimes measured with different questions. This might partially explain the differences in phenotypic correlations between studies. Some phenotypes with important differences in measurement between surveys are Cholesterol, Hearing difficulty, Osteoarthritis, and Type I and II Diabetes. 

\subsection{Childhood SES}
\label{subsec:SESmeasures} 

We measure the Socioeconomic Status (SES) of individuals in their childhood through an estimate of parental SES. We estimate a General structural equation model, with observed and latent variables. For the observed variables we used:
\begin{itemize}
\item Mother's years of education
\item Father's years of education
\item Father's occupation 
\end{itemize}
These variables exist in each of the HRS, WLS, and ELSA, with only minor differences.
Father's occupation variables are defined as 0 for ``Other'', 1 for ``Farmers'', 2 for ``Services + Operators + Admin/Clerical'', 3 for ``Sales'', 4 for ``Management'', and 5 for ``Professional''. These categories were assigned based on an analysis of income and education levels in the 3 studies. ELSA does not include the option ``Farmer'', and therefore that category is excluded only for ELSA.
The years of education reported by the father and mother vary between surveys. In HRS and ELSA the minimum number of years of parental education reported is 0 and the maximum is 14 for ELSA and 17 or more for HRS. In WLS the minimum is 7 years and the maximum 18.
From the general structural equation model, we calculate a standardized SES latent factor.

The distribution of the SES factor for ELSA is much bigger than for the other two studies due to the presence of a higher rate of respondents that reported 0 years of maternal and/or paternal education. We therefore imputed 7 years of education for such respondents and included a dummy variable to identify these individuals in the structural model to obtain a narrower and more comparable distribution of the SES factor.

\subsection{Polygenic Index}
PGIs were constructed using GWAS results that leave HRS, WLS, and ELSA out for reasons of statistical independence. We selected GWASs from the NHGRI-EBI GWAS Catalog \citep{buniello2019nhgri} and the GWAS Atlas \citep{watanabe2019global}, restricting to phenotypes available in all three datasets. We applied criteria for sufficient statistical power, including GWAS sample size, whether a superior GWAS was available for the same phenotype, the number of genome-wide significant hits, SNP-based heritability, and phenotype prevalence in the population, selecting the most relevant phenotype within each domain.

We generated two types of PGI for each phenotype: Full SNPs PGI and Top SNPs PGI. PGIs were constructed using PLINK as weighted sums of allele counts using GWAS effect sizes as weights, without linkage disequilibrium pruning or clumping. The Full SNPs PGI aggregates all available SNPs, capturing genetic associations across the whole genome (\textit{p}-value $\leq 1$). The Top SNPs PGI aggregates only SNPs that reach genome-wide significance in the GWAS (\textit{p}-value $\leq 5\times10^{-8}$). To account for multiple hypothesis testing across approximately one million independent SNPs in the human genome (adjacent SNPs are often in linkage disequilibrium, i.e., inherited together), the commonly used threshold for genome-wide statistical significance is $\textit{p} < 5 \times 10^{-8}$ (i.e., $0.05/1{,}000{,}000$).

Genetic quality control steps for the target datasets (HRS, WLS, and ELSA) were conducted following guidelines recommended in the literature \citep{marees2018tutorial}.

\subsection{Statistical Methods}

We estimate the relationship between phenotypes, polygenic indices (PGIs), socioeconomic status (SES), and the interaction between SES and PGI using pooled data from HRS, ELSA, and WLS. The model is estimated by OLS on the pooled sample and includes dataset fixed effects, as well as controls for age, sex, and the first 10 genetic principal components. Prior to estimation, phenotypes are standardized to have mean zero and variance one. The Full SNPs PGI and the SES factor are also standardized to have mean zero and variance one. As a robustness check, we also estimated the models separately for each dataset and using the top SNPs PGI aggregating the genome-wide significant SNPs. PGIs restricted to genome-wide significant SNPs generally lacked power and are therefore not presented.

\bibliography{gxses}

@incollection{cunha2006interpreting,
  title={Interpreting the evidence on life cycle skill formation},
  author={Cunha, Flavio and Heckman, James J and Lochner, Lance and Masterov, Dimitriy V},
  booktitle={Handbook of the Economics of Education},
  volume={1},
  pages={697--812},
  year={2006},
  publisher={Elsevier}
}

@article{galobardes2008association,
  title={Is the association between childhood socioeconomic circumstances and cause-specific mortality established? Update of a systematic review},
  author={Galobardes, Bruna and Lynch, John W and Smith, G Davey},
  journal={Journal of Epidemiology \& Community Health},
  volume={62},
  number={5},
  pages={387--390},
  year={2008},
  publisher={BMJ Publishing Group Ltd}
}

@article{buniello2019nhgri,
  title={The NHGRI-EBI GWAS Catalog of published genome-wide association studies, targeted arrays and summary statistics 2019},
  author={Buniello, Annalisa and MacArthur, Jacqueline A L and Cerezo, Maria and Harris, Laura W and Hayhurst, James and Malangone, Cinzia and McMahon, Aoife and Morales, Joannella and Mountjoy, Edward and Sollis, Elliot and others},
  journal={Nucleic Acids Research},
  volume={47},
  number={D1},
  pages={D1005--D1012},
  year={2019},
  publisher={Oxford University Press}
}

@article{hill2016molecular,
  title={Molecular genetic contributions to social deprivation and household income in UK Biobank},
  author={Hill, W David and Hagenaars, Saskia P and Marioni, Riccardo E and Harris, Sarah E and Liewald, David CM and Davies, Gail and Okbay, Aysu and McIntosh, Andrew M and Gale, Catharine R and Deary, Ian J},
  journal={Current Biology},
  volume={26},
  number={22},
  pages={3083--3089},
  year={2016},
  publisher={Elsevier}
}

@article{lee2018gene,
  title={Gene discovery and polygenic prediction from a genome-wide association study of educational attainment in 1.1 million individuals},
  author={Lee, James J and Wedow, Robbee and Okbay, Aysu and Kong, Edward and Maghzian, Omeed and Zacher, Meghan and Nguyen-Viet, Tuan Anh and Bowers, Peter and Sidorenko, Julia and Linn{\'e}r, Richard Karlsson and others},
  journal={Nature Genetics},
  volume={50},
  number={8},
  pages={1112--1121},
  year={2018},
  publisher={Nature Publishing Group}
}

@article{polderman2015meta,
  title={Meta-analysis of the heritability of human traits based on fifty years of twin studies},
  author={Polderman, Tinca JC and Benyamin, Beben and De Leeuw, Christiaan A and Sullivan, Patrick F and Van Bochoven, Arjen and Visscher, Peter M and Posthuma, Danielle},
  journal={Nature Genetics},
  volume={47},
  number={7},
  pages={702--709},
  year={2015},
  publisher={Nature Publishing Group}
}

@article{savage2018genome,
  title={Genome-wide association meta-analysis in 269,867 individuals identifies new genetic and functional links to intelligence},
  author={Savage, Jeanne E and Jansen, Philip R and Stringer, Sven and Watanabe, Kyoko and Bryois, Julien and De Leeuw, Christiaan A and Nagel, Mats and Awasthi, Swapnil and Barr, Peter B and Coleman, Jonathan RI and others},
  journal={Nature genetics},
  volume={50},
  number={7},
  pages={912--919},
  year={2018},
  publisher={Nature Publishing Group}
}

@article{turkheimer2000three,
  title={Three laws of behavior genetics and what they mean},
  author={Turkheimer, Eric},
  journal={Current Directions in Psychological Science},
  volume={9},
  number={5},
  pages={160--164},
  year={2000},
  publisher={SAGE Publications Sage CA: Los Angeles, CA}
}

@techreport{Lachman1997,
author = {Lachman, Margie E. and Weaver, Susanne L.},
institution = {Psycology Department, Brandeis University},
pages = {9},
title = {{The Midlife Development Inventory (MIDI) personality scales: Scale construction and scoring}},
url = {https://www.brandeis.edu/psychology/lachman/pdfs/midi-personality-scales.pdf},
year = {1997}
}

@techreport{Smith2017,
author = {Smith, Jacqui and Ryan, Lindsay and Sonnega, Amanda and Weir, David},
institution = {Survey Research Center Institute for Social Research University of Michigan Ann Arbor, Michigan},
title = {{Psychosocial and lifestyle questionnaire 2006-2016 documentation report core section LB}},
url = {https://hrs.isr.umich.edu/sites/default/files/biblio/HRS 2006-2016 SAQ Documentation{\_}07.06.17.pdf},
year = {2017}
}

@article{Kimball2008,
author = {Kimball, Miles S. and Sahm, Claudia R. and Shapiro, Matthew D.},
doi = {10.1198/016214508000000139},
issn = {01621459},
journal = {Journal of the American Statistical Association},
number = {483},
pages = {1028--1038},
title = {{Imputing risk tolerance from survey responses}},
volume = {103},
year = {2008}
}

@article{VandeVelde2009,
author = {{Van de Velde}, S and Levecque, K and Bracke, P},
doi = {10.1186/0778-7367-67-1-15},
issn = {2049-3258},
journal = {Archives of Public Health},
number = {1},
pages = {15},
title = {{Measurement equivalence of the CES-D 8 in the general population in Belgium: A gender perspective}},
url = {https://archpublichealth.biomedcentral.com/articles/10.1186/0778-7367-67-1-15},
volume = {67},
year = {2009}
}

@article{Karim2015,
author = {Karim, Jahanvash and Weisz, Robert and Bibi, Zainab and ur Rehman, Shafiq},
doi = {10.1007/s12144-014-9281-y},
issn = {19364733},
journal = {Current Psychology},
number = {4},
pages = {681--692},
publisher = {Springer New York LLC},
title = {{Validation of the eight-item Center for Epidemiologic Studies Depression Scale (CES-D) among older adults}},
volume = {34},
year = {2015}
}

@article{Hughes2004,
  title={A short scale for measuring loneliness in large surveys: Results from two population-based studies},
  author={Hughes, Mary Elizabeth and Waite, Linda J and Hawkley, Louise C and Cacioppo, John T},
  journal={Research on Aging},
  volume={26},
  number={6},
  pages={655--672},
  year={2004},
  publisher={Sage Publications Sage CA: Thousand Oaks, CA}
}

@techreport{Steffick2000,
author = {Steffick, Diane E and Wallace, Robert B and Regula, A and Mary, Herzog and Ofstedal, Beth and Steffick, Diane and Fonda, Stephanie and Langa, Ken},
title = {{HRS/AHEAD cocumentation report documentation of affective functioning measures in the Health and Retirement Study}},
institution = {University of Michigan},
year = {2000}
}

@article{Steptoe2013,
author = {Steptoe, A. and Breeze, E. and Banks, J. and Nazroo, J.},
doi = {10.1093/ije/dys168},
issn = {0300-5771},
journal = {International Journal of Epidemiology},
number = {6},
pages = {1640--1648},
title = {{Cohort profile: The English Longitudinal Study of Ageing}},
url = {https://academic.oup.com/ije/article-lookup/doi/10.1093/ije/dys168},
volume = {42},
year = {2013}
}

@article{Diener1985,
author = {Diener, Ed and Emmons, Robert A. and Larsem, Randy J. and Griffin, Sharon},
doi = {10.1207/s15327752jpa4901_13},
issn = {15327752},
journal = {Journal of Personality Assessment},
number = {1},
pages = {71--75},
pmid = {16367493},
title = {{The satisfaction with life scale}},
volume = {49},
year = {1985}
}

@article{Gale2014,
author = {Gale, Catharine R. and Allerhand, Michael and Sayer, Avan Aihie and Cooper, Cyrus and Deary, Ian J.},
doi = {10.1007/s11357-014-9682-8},
issn = {15744647},
journal = {Age},
number ={1},
pages={9682},
publisher = {Kluwer Academic Publishers},
title = {{The dynamic relationship between cognitive function and walking speed: The English Longitudinal Study of Ageing}},
volume = {36},
year = {2014}
}

@article{Stephan2018,
author = {Stephan, Yannick and Sutin, Angelina R. and Kornadt, Anna and Caudroit, Johan and Terracciano, Antonio},
doi = {10.1016/j.intell.2018.06.006},
issn = {01602896},
journal = {Intelligence},
pages = {195--199},
publisher = {Elsevier Ltd},
title = {{Higher IQ in adolescence is related to a younger subjective age in later life: Findings from the Wisconsin Longitudinal Study}},
volume = {69},
number={1},
year = {2018}
}

@techreport{WLSScale,
author = {{Center for Demography of Health and Aging}},
institution = {University of Wisconsin},
title = {{Documentation of scales in Wisconsin Longitudinal Study}},
year = {1992}
}

@techreport{Savard1957,
author = {Savard, Joe R and Pearce, Noah C and Parks, Ross A and Bell, Nick A and Stark, Marie E and Potvin, Annabelle R},
title = {{Wisconsin Longitudinal Study user's guide}},
institution = {University of Wisconsin},
url = {http://www.ssc.wisc.edu/wlsresearch},
year = {1957}
}

@article{marees2018tutorial,
  title={A tutorial on conducting genome-wide association studies: Quality control and statistical analysis},
  author={Marees, Andries T and de Kluiver, Hilde and Stringer, Sven and Vorspan, Florence and Curis, Emmanuel and Marie-Claire, Cynthia and Derks, Eske M},
  journal={International Journal of Methods in Psychiatric Research},
  volume={27},
  number={2},
  pages={e1608},
  year={2018},
  publisher={Wiley Online Library}
}

@article{lin2010new,
  title={A new statistic to evaluate imputation reliability},
  author={Lin, Peng and Hartz, Sarah M and Zhang, Zhehao and Saccone, Scott F and Wang, Jia and Tischfield, Jay A and Edenberg, Howard J and Kramer, John R and Goate, Alison M and Bierut, Laura J and others},
  journal={PloS ONE},
  volume={5},
  number={3},
  year={2010},
  publisher={Public Library of Science}
}

@article{akaneya2010ephrin,
  title={Ephrin-A5 and EphA5 interaction induces synaptogenesis during early hippocampal development},
  author={Akaneya, Yukio and Sohya, Kazuhiro and Kitamura, Akihiko and Kimura, Fumitaka and Washburn, Chris and Zhou, Renping and Ninan, Ipe and Tsumoto, Tadaharu and Ziff, Edward B},
  journal={PloS ONE},
  volume={5},
  number={8},
  pages={e9697},
  year={2010},
  publisher={Public Library of Science}
}

@article{watanabe2019global,
  title={A global overview of pleiotropy and genetic architecture in complex traits},
  author={Watanabe, Kyoko and Stringer, Sven and Frei, Oleksandr and Mirkov, Ma{\v{s}}a Umi{\'c}evi{\'c} and de Leeuw, Christiaan and Polderman, Tinca JC and van der Sluis, Sophie and Andreassen, Ole A and Neale, Benjamin M and Posthuma, Danielle},
  journal={Nature Genetics},
  volume={51},
  number={9},
  pages={1339--1348},
  year={2019},
  publisher={Nature Publishing Group}
}

@article{weir2012quality,
  title={Quality control report for genotypic data},
  author={Weir, D and Faul, JD and Kardia, SL and Smith, JD and Doheny, KF and Romm, J and others},
  journal={Health and Retirement Study},
  year={2012}
}

@article{10002015global,
  title={A global reference for human genetic variation},
  author={{1000 Genomes Project Consortium} and others},
  journal={Nature},
  volume={526},
  number={7571},
  pages={68--74},
  year={2015},
  publisher={Nature Publishing Group}
}

@article{Sonnega2014cohort,
author = {Sonnega, Amanda and Faul, Jessica D and Ofstedal, M B and Langa, K M and Phillips, John W and Weir, David R},
doi = {10.1093/ije/dyu067},
issn = {0300-5771},
journal = {International Journal of Epidemiology},
number = {2},
pages = {576--585},
pmid = {24671021},
title = {{Cohort profile: the Health and Retirement Study (HRS)}},
url = {http://www.ncbi.nlm.nih.gov/pubmed/24671021 http://www.pubmedcentral.nih.gov/articlerender.fcgi?artid=PMC3997380 https://academic.oup.com/ije/article-lookup/doi/10.1093/ije/dyu067},
volume = {43},
year = {2014}
}

@article{Becker1986,
author = {Becker, Gary S and Tomes, Nigel},
issn = {0734-306X},
journal = {Journal of Labor Economics},
number = {3 Part 2},
pages = {1--47},
pmid = {12146356},
title = {{Human capital and the rise and fall of families.}},
url = {http://www.jstor.org/stable/2534952},
volume = {4},
year = {1986}
}

@article{Turkheimer2003,
author = {Turkheimer, Eric N and Haley, Andreana and Waldron, Mary and D'Onofrio, Brian M and Gottesman, Irving I},
doi = {10.1046/j.0956-7976.2003.psci_1475.x},
issn = {0956-7976},
journal = {Psychological Science},
language = {en},
number = {6},
pages = {623--628},
pmid = {14629696},
publisher = {SAGE Publications},
title = {{Socioeconomic status modifies heritability of IQ in young children}},
url = {http://pss.sagepub.com/content/14/6/623.full},
volume = {14},
year = {2003}
}

@article{Black2005,
author = {Black, Sandra E and Devereux, Paul J and Salvanes, Kjell G},
doi = {10.1257/0002828053828635},
issn = {0002-8282},
journal = {American Economic Review},
number = {1},
pages = {437--449},
title = {{Why the apple doesn't fall far: Understanding intergenerational transmission of human capital}},
url = {http://pubs.aeaweb.org/doi/10.1257/0002828053828635},
volume = {95},
year = {2005}
}

@article{Chetty2020,
author = {Chetty, Raj and Friedman, John N and Saez, Emmanuel and Turner, Nicholas and Yagan, Danny},
doi = {10.1093/qje/qjaa005},
issn = {0033-5533},
journal = {Quarterly Journal of Economics},
number = {3},
pages = {1567--1633},
title = {Income segregation and intergenerational mobility across colleges in the United States},
url = {https://academic.oup.com/qje/article/135/3/1567/5741707},
volume = {135},
year = {2020}
}

@article{miao2025pigeon,
  title={PIGEON: a statistical framework for estimating gene--environment interaction for polygenic traits},
  author={Miao, Jiacheng and Song, Gefei and Wu, Yixuan and Hu, Jiaxin and Wu, Yuchang and Basu, Shubhashrita and Andrews, James S and Schaumberg, Katherine and Fletcher, Jason M and Schmitz, Lauren L and others},
  journal={Nature human behaviour},
  volume={9},
  number={8},
  pages={1654--1668},
  year={2025},
  publisher={Nature Publishing Group UK London}
}

@article{hout1993persistence,
  title={The persistence of classes in post-industrial societies},
  author={Hout, Mike and Brooks, Clem and Manza, Jeff},
  journal={International Sociology},
  volume={8},
  number={3},
  pages={259--277},
  year={1993},
  publisher={Sage Publications}
}

@article{siponen2011children,
  title={Children's health and parental socioeconomic factors: a population-based survey in Finland},
  author={Siponen, Sanna M and Ahonen, Riitta S and Savolainen, Piia H and H{\"a}meen-Anttila, Katri P},
  journal={BMC Public Health},
  volume={11},
  number={1},
  pages={1--8},
  year={2011},
  publisher={BioMed Central}
}

@article{GWASCatalog_2013,
    author = {Welter, Danielle and MacArthur, Jacqueline and Morales, Joannella and Burdett, Tony and Hall, Peggy and Junkins, Heather and Klemm, Alan and Flicek, Paul and Manolio, Teri and Hindorff, Lucia and Parkinson, Helen},
    title = "{The NHGRI GWAS Catalog, a curated resource of SNP-trait associations}",
    journal = {Nucleic Acids Research},
    volume = {42},
    number = {D1},
    pages = {D1001-D1006},
    year = {2013},
    issn = {0305-1048},
    doi = {10.1093/nar/gkt1229},
    url = {https://doi.org/10.1093/nar/gkt1229},
    eprint = {https://academic.oup.com/nar/article-pdf/42/D1/D1001/3646160/gkt1229.pdf},
}

@article{Tyrrell2017,
    author = {Tyrrell, Jessica and Wood, Andrew R and Ames, Ryan M and Yaghootkar, Hanieh and Beaumont, Robin N and Jones, Samuel E and Tuke, Marcus A and Ruth, Katherine S and Freathy, Rachel M and Davey Smith, George and Joost, StÃ©phane and Guessous, Idris and Murray, Anna and Strachan, David P and Kutalik, ZoltÃ¡n and Weedon, Michael N and Frayling, Timothy M},
    title = "{Gene-obesogenic environment interactions in the UK Biobank study}",
    journal = {International Journal of Epidemiology},
    volume = {46},
    number = {2},
    pages = {559-575},
    year = {2017},
    issn = {0300-5771},
    doi = {10.1093/ije/dyw337},
    url = {https://doi.org/10.1093/ije/dyw337},
    eprint = {https://academic.oup.com/ije/article-pdf/46/2/559/24172372/dyw337.pdf},
}

@article{Haldane1946,
author = {Haldane, John Burdon Sanderson},
journal = {Annals of Eugenics},
number = {3},
pages = {197--205},
pmid = {20282564},
title = {{The interaction of nature and nurture.}},
url = {http://www.ncbi.nlm.nih.gov/pubmed/20282564},
volume = {13},
year = {1946}
}

@article {Domingue_2020B,
author = {Benjamin W. Domingue and Sam Trejo and Emma Armstrong-Carter and Elliot M. Tucker-Drob },
title = {Interactions between polygenic scores and environments: Methodological and conceptual challenges},
journal = {Sociological Science},
volume = {7},
number={1},
issn = {2330-6696},
url = {http://dx.doi.org/10.15195/v7.a19},
doi = {10.15195/v7.a19},
pages = {465--486},
year = {2020},
}

@article{pasman2019systematic,
  title={Systematic review of polygenic gene--environment interaction in tobacco, alcohol, and cannabis use},
  author={Pasman, Jo{\"e}lle A and Verweij, Karin JH and Vink, Jacqueline M},
  journal={Behavior Genetics},
  volume={49},
  number={4},
  pages={349--365},
  year={2019},
  publisher={Springer}
}

@article{yengo2018meta,
  title={Meta-analysis of genome-wide association studies for height and body mass index in \~{}700000 individuals of European ancestry},
  author={Yengo, Loic and Sidorenko, Julia and Kemper, Kathryn E and Zheng, Zhili and Wood, Andrew R and Weedon, Michael N and Frayling, Timothy M and Hirschhorn, Joel and Yang, Jian and Visscher, Peter M and others},
  journal={Human Molecular Genetics},
  volume={27},
  number={20},
  pages={3641--3649},
  year={2018},
  publisher={Oxford University Press}
}

@article{schmitz2021impact,
  title={The impact of late-career job loss and genetic risk on body mass index: Evidence from variance polygenic scores},
  author={Schmitz, Lauren L and Goodwin, Julia and Miao, Jiacheng and Lu, Qiongshi and Conley, Dalton},
  journal={Scientific Reports},
  volume={11},
  number={1},
  pages={1--15},
  year={2021},
  publisher={Nature Publishing Group}
}

@article{Schmitz2016,
author = {Schmitz, Lauren and Conley, Dalton},
doi = {10.1007/s10519-015-9739-1},
issn = {0001-8244},
journal = {Behavior Genetics},
number = {1},
pages = {43--58},
publisher = {Springer US},
title = {{The long-term consequences of Vietnam-era conscription and genotype on smoking behavior and health}},
url = {http://link.springer.com/10.1007/s10519-015-9739-1},
volume = {46},
year = {2016}
}

@article{Domingue2015,
  title={Cohort effects in the genetic influence on smoking},
  author={Domingue, Benjamin W and Conley, Dalton and Fletcher, Jason and Boardman, Jason D},
  journal={Behavior Genetics},
  volume={46},
  number={1},
  pages={31--42},
  year={2016},
  publisher={Springer}
}

@article{Meyers2013,
author = {Meyers, Jacquelyn L and Cerd{\'{a}}, Magdalena and Galea, Sandro and Keyes, Katherine M and Aiello, Allison E and Uddin, Monica and Wildman, Derek E and Koenen, Karestan C},
doi = {10.1038/tp.2013.63},
issn = {2158-3188},
journal = {Translational Psychiatry},
number = {8},
pages = {e290},
pmid = {23942621},
publisher = {Macmillan Publishers Limited},
shorttitle = {Transl Psychiatry},
title = {{Interaction between polygenic risk for cigarette use and environmental exposures in the Detroit neighborhood health study}},
url = {http://dx.doi.org/10.1038/tp.2013.63 http://www.nature.com/doifinder/10.1038/tp.2013.63},
volume = {3},
year = {2013}
}

@article{Fletcher2012,
author = {Fletcher, Jason M},
doi = {10.1371/journal.pone.0050576},
issn = {1932-6203},
journal = {PloS ONE},
number = {12},
pages = {e50576},
pmid = {23227187},
title = {{Why have tobacco control policies stalled? Using genetic moderation to examine policy impacts.}},
url = {http://www.pubmedcentral.nih.gov/articlerender.fcgi?artid=3515624{\&}tool=pmcentrez{\&}rendertype=abstract},
volume = {7},
year = {2012}
}

@article{Treur2017,
  title={Testing familial transmission of smoking with two different research designs},
  author={Treur, Jorien L and Verweij, Karin JH and Abdellaoui, Abdel and Fedko, Iryna O and de Zeeuw, Eveline L and Ehli, Erik A and Davies, Gareth E and Hottenga, Jouke-Jan and Willemsen, Gonneke and Boomsma, Dorret I and others},
  journal={Nicotine and Tobacco Research},
  volume={20},
  number={7},
  pages={836--842},
  year={2018},
  publisher={Oxford University Press US}
}

@article{GSCAN2019gwas,
author = {Liu, Mengzhen and Jiang, Yu and Wedow, Robbee and Li, Yue and Brazel, David M and Chen, Fang and Datta, Gargi and ... and Bierut, Laura J and Hveem, Kristian and Lee, James J and Munafo, Marcus R and Saccone, Nancy L and Willer, Cristen J and Cornelis, Marilyn C and David, Sean P and Hinds, David A and Jorgenson, Eric and Kaprio, Jaakko and Stitzel, Jerry A and Stefansson, Kari and Thorgeirsson, Thorgeir E and Abecasis, Goncalo R and Liu, Dajiang J and Vrieze, Scott},
doi = {10.1038/s41588-018-0307-5},
issn = {1061-4036},
journal = {Nature Genetics},
title = {{Association studies of up to 1.2 million individuals yield new insights into the genetic etiology of tobacco and alcohol use}},
url = {http://www.nature.com/articles/s41588-018-0307-5},
year = {2019}
}

@article{scarr1971race,
  title={Race, social class, and IQ},
  author={Scarr-Salapatek, Sandra},
  journal={Science},
  volume={174},
  number={4016},
  pages={1285--1295},
  year={1971},
  publisher={JSTOR}
}

@article{rowe1999genetic,
  title={Genetic and environmental influences on vocabulary IQ: Parental education level as moderator},
  author={Rowe, David C and Jacobson, Kristen C and Van den Oord, Edwin JCG},
  journal={Child development},
  volume={70},
  number={5},
  pages={1151--1162},
  year={1999},
  publisher={Wiley Online Library}
}

@article{barker1993fetal,
  title={Fetal nutrition and cardiovascular disease in adult life},
  author={Barker, David JP and Godfrey, Keith M and Gluckman, Peter D and Harding, Jane E and Owens, Julie A and Robinson, Jeffrey S},
  journal={Lancet},
  volume={341},
  number={8850},
  pages={938--941},
  year={1993},
  publisher={Elsevier}
}

@incollection{assary2018gene,
  title={Gene-environment interaction and psychiatric disorders: Review and future directions},
  author={Assary, Elham and Vincent, John Paul and Keers, Robert and Pluess, Michael},
  booktitle={Seminars in cell \& developmental biology},
  volume={77},
  pages={133--143},
  year={2018},
  publisher={Elsevier}
}

@article{bulik2015ld,
  title={LD Score regression distinguishes confounding from polygenicity in genome-wide association studies},
  author={Bulik-Sullivan, Brendan K and Loh, Po-Ru and Finucane, Hilary K and Ripke, Stephan and Yang, Jian and Patterson, Nick and Daly, Mark J and Price, Alkes L and Neale, Benjamin M},
  journal={Nature Genetics},
  volume={47},
  number={3},
  pages={291--295},
  year={2015},
  publisher={Nature Publishing Group}
}

@article{howard2019genome,
  title={Genome-wide meta-analysis of depression identifies 102 independent variants and highlights the importance of the prefrontal brain regions},
  author={Howard, David M and Adams, Mark J and Clarke, Toni-Kim and Hafferty, Jonathan D and Gibson, Jude and Shirali, Masoud and Coleman, Jonathan RI and Hagenaars, Saskia P and Ward, Joey and Wigmore, Eleanor M and others},
  journal={Nature Neuroscience},
  volume={22},
  number={3},
  pages={343--352},
  year={2019},
  publisher={Nature Publishing Group}
}

@article{bulik2015atlas,
  title={An atlas of genetic correlations across human diseases and traits},
  author={Bulik-Sullivan, Brendan and Finucane, Hilary K and Anttila, Verneri and Gusev, Alexander and Day, Felix R and Loh, Po-Ru and Duncan, Laramie and Perry, John RB and Patterson, Nick and Robinson, Elise B and others},
  journal={Nature Genetics},
  volume={47},
  number={11},
  pages={1236--1241},
  year={2015},
  publisher={Nature Publishing Group}
}

@article{herd2014cohort,
  title={Cohort profile: Wisconsin Longitudinal Study (WLS)},
  author={Herd, Pamela and Carr, Deborah and Roan, Carol},
  journal={International Journal of Epidemiology},
  volume={43},
  number={1},
  pages={34--41},
  year={2014},
  publisher={Oxford University Press}
}

@article{mcallister2017current,
  title={Current challenges and new opportunities for gene-environment interaction studies of complex diseases},
  author={McAllister, Kimberly and Mechanic, Leah E and Amos, Christopher and Aschard, Hugues and Blair, Ian A and Chatterjee, Nilanjan and Conti, David and Gauderman, W James and Hsu, Li and Hutter, Carolyn M and others},
  journal={American Journal of Epidemiology},
  volume={186},
  number={7},
  pages={753--761},
  year={2017},
  publisher={Oxford University Press}
}

@article{plomin1977genotype,
  title={Genotype-environment interaction and correlation in the analysis of human behavior.},
  author={Plomin, Robert and DeFries, John C and Loehlin, John C},
  journal={Psychological Bulletin},
  volume={84},
  number={2},
  pages={309},
  year={1977},
  publisher={American Psychological Association}
}

@article{barcellos2018education,
  title={Education can reduce health differences related to genetic risk of obesity},
  author={Barcellos, Silvia H and Carvalho, Leandro S and Turley, Patrick},
  journal={Proceedings of the National Academy of Sciences},
  volume={115},
  number={42},
  pages={E9765--E9772},
  year={2018},
  publisher={National Acad Sciences}
}

@article{amin2017gene,
  title={Gene-environment interactions between education and body mass: Evidence from the {UK} and {Finland}},
  author={Amin, Vikesh and B{\"o}ckerman, Petri and Viinikainen, Jutta and Smart, Melissa C and Bao, Yanchun and Kumari, Meena and Pitk{\"a}nen, Niina and Lehtim{\"a}ki, Terho and Raitakari, Olli and Pehkonen, Jaakko},
  journal={Social Science \& Medicine},
  volume={195},
  number={1},
  pages={12--16},
  year={2017},
  publisher={Elsevier}
}

@article{hoang2023heterogenous,
  title={Heterogenous trajectories in physical, mental and cognitive health among older Americans: Roles of genetics and life course contextual factors},
  author={Hoang, Cung Truong and Amin, Vikesh and Behrman, Jere R and Kohler, Hans-Peter and Kohler, Illiana V},
  journal={SSM-Population Health},
  volume={23},
  number={1},
  pages={101448},
  year={2023},
  publisher={Elsevier}
}

@article{bierut2018childhood,
  title={Challenges in studying the interplay of genes and environment. A study of childhood financial distress moderating genetic predisposition for peak smoking},
  author={Bierut, Laura and Biroli, Pietro and Galama, Titus J and Thom, Kevin},
  journal={Journal of Economic Psychology},
  volume={98},
  number={1},
  pages={102636},
  year={2023},
  publisher={Elsevier}
}

@article{peplow2014social,
  title={Social sciences suffer from severe publication bias},
  author={Peplow, Mark},
  journal={Nature},
  volume={10},
  year={2014}
}

@article{franco2014publication,
  title={Publication bias in the social sciences: Unlocking the file drawer},
  author={Franco, Annie and Malhotra, Neil and Simonovits, Gabor},
  journal={Science},
  volume={345},
  number={6203},
  pages={1502--1505},
  year={2014},
  publisher={American Association for the Advancement of Science}
}

@article{sterling1959publication,
  title={Publication decisions and their possible effects on inferences drawn from tests of significance—or vice versa},
  author={Sterling, Theodore D},
  journal={Journal of the American statistical association},
  volume={54},
  number={285},
  pages={30--34},
  year={1959},
  publisher={Taylor \& Francis}
}

@article{teicher2016effects,
  title={The effects of childhood maltreatment on brain structure, function and connectivity},
  author={Teicher, Martin H and Samson, Jacqueline A and Anderson, Carl M and Ohashi, Kyoko},
  journal={Nature Reviews Neuroscience},
  volume={17},
  number={10},
  pages={652--666},
  year={2016},
  publisher={Nature Publishing Group UK London}
}

@article{cowell2015childhood,
  title={Childhood maltreatment and its effect on neurocognitive functioning: Timing and chronicity matter},
  author={Cowell, Raquel A and Cicchetti, Dante and Rogosch, Fred A and Toth, Sheree L},
  journal={Development and Psychopathology},
  volume={27},
  number={2},
  pages={521--533},
  year={2015},
  publisher={Cambridge University Press}
}

@article{teicher2016annual,
  title={Annual research review: Enduring neurobiological effects of childhood abuse and neglect},
  author={Teicher, Martin H and Samson, Jacqueline A},
  journal={Journal of Child Psychology and Psychiatry},
  volume={57},
  number={3},
  pages={241--266},
  year={2016},
  publisher={Wiley Online Library}
}

@article{norman2012long,
  title={The long-term health consequences of child physical abuse, emotional abuse, and neglect: A systematic review and meta-analysis},
  author={Norman, Rosana E and Byambaa, Munkhtsetseg and De, Rumna and Butchart, Alexander and Scott, James and Vos, Theo},
  journal={PLoS Medicine},
  volume={9},
  number={11},
  pages={e1001349},
  year={2012},
  publisher={Public Library of Science San Francisco, USA}
}

@article{gardner2019association,
  title={The association between five forms of child maltreatment and depressive and anxiety disorders: A systematic review and meta-analysis},
  author={Gardner, Madeleine J and Thomas, Hannah J and Erskine, Holly E},
  journal={Child Abuse \& Neglect},
  volume={96},
  number={1},
  pages={104082},
  year={2019},
  publisher={Elsevier}
}

@article{d2022childhood,
  title={Childhood abuse and neglect, and mortality risk in adulthood: A systematic review and meta-analysis},
  author={D'arcy-Bewick, Sinead and Terracciano, Antonio and Turiano, Nicholas and Sutin, Angelina R and Long, Roisin and O'S{\'u}illeabh{\'a}in, P{\'a}raic S},
  journal={Child Abuse \& Neglect},
  volume={134},
  number={1},
  pages={105922},
  year={2022},
  publisher={Elsevier}
}

@book{Harden2021book,
  title = {The genetic lottery: Why {{DNA}} matters for social equality},
  author = {Harden, Kathryn Paige},
  year = {2021},
  publisher = {Princeton University Press},
  urldate = {2021-08-12},
  isbn = {978-0-691-19080-8}
}

@article{ghirardi2024Interaction,
  title = {Interaction of family {{SES}} with children's genetic propensity for cognitive and noncognitive skills: {{No}} evidence of the {{Scarr-Rowe}} hypothesis for educational outcomes},
  shorttitle = {Interaction of Family {{SES}} with Children's Genetic Propensity for Cognitive and Noncognitive Skills},
  author = {Ghirardi, Gaia and {Gil-Hern{\'a}ndez}, Carlos J. and Bernardi, Fabrizio and {van Bergen}, Elsje and Demange, Perline},
  year = {2024},
  journal = {Research in Social Stratification and Mobility},
  volume = {92},
  number={1},
  pages = {100960},
  issn = {0276-5624},
  doi = {10.1016/j.rssm.2024.100960},
  urldate = {2025-04-11}
}

@article{ghirardi2025Compensating,
  title = {Compensating or boosting genetic propensities? {{Gene-family}} socioeconomic status interactions by educational outcome selectivity},
  shorttitle = {Compensating or Boosting Genetic Propensities?},
  author = {Ghirardi, Gaia and Bernardi, Fabrizio},
  year = {2025},
  journal = {Social Science Research},
  volume = {129},
  number={1},
  pages = {103174},
  issn = {0049-089X},
  doi = {10.1016/j.ssresearch.2025.103174},
  urldate = {2025-04-11},
}

@article{papageorge2020genes,
  title={Genes, education, and labor market outcomes: evidence from the Health and Retirement Study},
  author={Papageorge, Nicholas W and Thom, Kevin},
  journal={Journal of the European Economic Association},
  volume={18},
  number={3},
  pages={1351--1399},
  year={2020},
  publisher={Oxford University Press}
}

@article{houmark2024nurture,
  title={The nurture of nature and the nature of nurture: How genes and investments interact in the formation of skills},
  author={Houmark, Mikkel Aagaard and Ronda, Victor and Rosholm, Michael},
  journal={American Economic Review},
  volume={114},
  number={2},
  pages={385--425},
  year={2024},
  publisher={American Economic Association 2014 Broadway, Suite 305, Nashville, TN 37203}
}

@article{harden2007genotype,
  title={Genotype by environment interaction in adolescents’ cognitive aptitude},
  author={Harden, K Paige and Turkheimer, Eric and Loehlin, John C},
  journal={Behavior Genetics},
  volume={37},
  number={2},
  pages={273--283},
  year={2007},
  publisher={Springer}
}

@article{tucker2016large,
  title={Large cross-national differences in gene$\times$ socioeconomic status interaction on intelligence},
  author={Tucker-Drob, Elliot M and Bates, Timothy C},
  journal={Psychological science},
  volume={27},
  number={2},
  pages={138--149},
  year={2016},
  publisher={Sage Publications Sage CA: Los Angeles, CA}
}

@article{johnson2022polygenic,
  title={Polygenic scores for plasticity: a new tool for studying gene--environment interplay},
  author={Johnson, Rebecca and Sotoudeh, Ramina and Conley, Dalton},
  journal={Demography},
  volume={59},
  number={3},
  pages={1045--1070},
  year={2022},
  publisher={Duke University Press}
}

@article{isungset2022social,
  title={Social and genetic associations with educational performance in a Scandinavian welfare state},
  author={Isungset, Martin A and Conley, Dalton and Zachrisson, Henrik D and Ystrom, Eivind and Havdahl, Alexandra and Nj{\o}lstad, P{\aa}l R and Lyngstad, Torkild Hovde},
  journal={Proceedings of the National Academy of Sciences},
  volume={119},
  number={25},
  pages={e2201869119},
  year={2022},
  publisher={National Academy of Sciences}
}

@article{becker2021resource,
  title={Resource profile and user guide of the Polygenic Index Repository},
  author={Becker, Joel and Burik, Casper AP and Goldman, Grant and Wang, Nancy and Jayashankar, Hariharan and Bennett, Michael and Belsky, Daniel W and Karlsson Linn{\'e}r, Richard and Ahlskog, Rafael and Kleinman, Aaron and others},
  journal={Nature Human Behaviour},
  volume={5},
  number={12},
  pages={1744--1758},
  year={2021},
  publisher={Nature Publishing Group UK London}
}

@article{biroli2026economics,
  title={The economics and econometrics of gene--environment interplay},
  author={Biroli, Pietro and Galama, Titus and Von Hinke, Stephanie and Van Kippersluis, Hans and Rietveld, Cornelius A and Thom, Kevin},
  journal={Review of Economic Studies},
  volume={93},
  number={1},
  pages={144--180},
  year={2025},
  publisher={Oxford University Press UK}
}

@article{Hanscombe2012,
  author  = {Hanscombe, Ken B. and Trzaskowski, Maciej and Haworth, Claire M. A. and Davis, Oliver S. P. and Dale, Philip S. and Plomin, Robert},
  title   = {Socioeconomic status (SES) and children's intelligence (IQ): In a UK-representative sample SES moderates the environmental, not genetic, effect on IQ},
  journal = {PLoS ONE},
  year    = {2012},
  volume  = {7},
  number  = {2},
  pages   = {e30320},
  doi     = {10.1371/journal.pone.0030320}
}

@book{BlauDuncan1967,
  author    = {Blau, Peter M. and Duncan, Otis Dudley},
  title     = {The American Occupational Structure},
  publisher = {Wiley},
  address   = {New York},
  year      = {1967}
}

@article{Mare2011,
  author  = {Mare, Robert D.},
  title   = {A multigenerational view of inequality},
  journal = {Demography},
  year    = {2011},
  volume  = {48},
  number  = {1},
  pages   = {1--23},
  doi     = {10.1007/s13524-011-0014-7}
}

@book{Shonkoff2000,
  author    = {Shonkoff, Jack P. and Phillips, Deborah A.},
  title     = {From neurons to neighborhoods: The science of early childhood development},
  publisher = {National Academy Press},
  address   = {Washington, DC},
  year      = {2000},
  doi       = {10.17226/9824}
}

@article{Heckman2006,
  author  = {Heckman, James J.},
  title   = {Skill formation and the economics of investing in disadvantaged children},
  journal = {Science},
  year    = {2006},
  volume  = {312},
  number  = {5782},
  pages   = {1900--1902},
  doi     = {10.1126/science.1128898}
}

@article{Solon1992,
  author  = {Solon, Gary},
  title   = {Intergenerational income mobility in the United States},
  journal = {American Economic Review},
  year    = {1992},
  volume  = {82},
  number  = {3},
  pages   = {393--408},
  doi     = {10.3386/w4098}
}

@article{Corak2013,
  author  = {Corak, Miles},
  title   = {Income inequality, equality of opportunity, and intergenerational mobility},
  journal = {Journal of Economic Perspectives},
  year    = {2013},
  volume  = {27},
  number  = {3},
  pages   = {79--102},
  doi     = {10.1257/jep.27.3.79}
}

@book{Jencks1972,
  author    = {Jencks, Christopher and Smith, Marshall and Acland, Henry and Bane, Mary Jo and Cohen, David and Gintis, Herbert and Heyns, Barbara and Michelson, Stephan},
  title     = {Inequality: A reassessment of the effect of family and schooling in America},
  publisher = {Basic Books},
  address   = {New York},
  year      = {1972}
}

@book{BourdieuPasseron1977,
  author    = {Bourdieu, Pierre and Passeron, Jean-Claude},
  title     = {Reproduction in Education, Society and Culture},
  publisher = {Sage},
  year      = {1977}
}

@article{Almond2018vol2,
  author  = {Almond, Douglas and Currie, Janet and Duque, Valentina},
  title   = {Childhood circumstances and adult outcomes: Act II},
  journal = {Journal of Economic Literature},
  year    = {2018},
  volume  = {56},
  number  = {4},
  pages   = {1360--1446},
  doi     = {10.1257/jel.20171164}
}

@incollection{Almond2011vol1,
  author    = {Almond, Douglas and Currie, Janet},
  title     = {Human capital development before age five},
  booktitle = {Handbook of Labor Economics},
  editor    = {Ashenfelter, Orley and Card, David},
  volume    = {4B},
  pages     = {1315--1486},
  publisher = {Elsevier},
  year      = {2011},
  doi       = {10.1016/S0169-7218(11)02413-0}
}

@article{Aizer2014intergen,
  author  = {Aizer, Anna and Currie, Janet},
  title   = {The intergenerational transmission of inequality: Maternal disadvantage and health at birth},
  journal = {Science},
  year    = {2014},
  volume  = {344},
  number  = {6186},
  pages   = {856--861},
  doi     = {10.1126/science.1251872}
}

@article{abdellaoui2023,
	author = {Abdellaoui, Abdel and Yengo, Loic and Verweij, Karin J. H. and Visscher, Peter M.},
	year = {2023},
	title = {15 years of {GWAS} discovery: {Realizing} the promise},
	volume = {110},
	doi = {10.1016/j.ajhg.2022.12.011},
	language = {English},
	number = {2},
	journal = {American Journal of Human Genetics},
	publisher = {Elsevier},
	pages = {179--194},
}

@article{van2023overcoming,
  title={Overcoming attenuation bias in regressions using polygenic indices},
  author={Van Kippersluis, Hans and Biroli, Pietro and Dias Pereira, Rita and Galama, Titus J and Von Hinke, Stephanie and Meddens, S Fleur W and Muslimova, Dilnoza and Slob, Eric AW and De Vlaming, Ronald and Rietveld, Cornelius A},
  journal={Nature Communications},
  volume={14},
  number={1},
  pages={4473},
  year={2023},
  publisher={Nature Publishing Group UK London}
}

@article{baselmans2019multivariate,
  title={Multivariate genome-wide analyses of the well-being spectrum},
  author={Baselmans, Bart ML and Jansen, Rick and Ip, Hill F and van Dongen, Jenny and Abdellaoui, Abdel and van de Weijer, Margot P and Bao, Yanchun and Smart, Melissa and Kumari, Meena and Willemsen, Gonneke and others},
  journal={Nature genetics},
  volume={51},
  number={3},
  pages={445--451},
  year={2019},
  publisher={Nature Publishing Group US New York}
}

@misc{NealeUKB2018,
  author = {Neale, Benjamin M. and et al.},
  title = {Rapid GWAS of thousands of phenotypes in the UK Biobank},
  year = {2018},
  howpublished = {\url{http://www.nealelab.is/uk-biobank}},
  note = {Neale Lab UK Biobank GWAS}
}

@article{mahajan2018fine,
  title={Fine-mapping type 2 diabetes loci to single-variant resolution using high-density imputation and islet-specific epigenome maps},
  author={Mahajan, Anubha and Taliun, Daniel and Thurner, Matthias and Robertson, Neil R and Torres, Jason M and Rayner, N William and Payne, Anthony J and Steinthorsdottir, Valgerdur and Scott, Robert A and Grarup, Niels and others},
  journal={Nature genetics},
  volume={50},
  number={11},
  pages={1505--1513},
  year={2018},
  publisher={Nature Publishing Group US New York}
}

@article{onengut2015fine,
  title={Fine mapping of type 1 diabetes susceptibility loci and evidence for colocalization of causal variants with lymphoid gene enhancers},
  author={Onengut-Gumuscu, Suna and Chen, Wei-Min and Burren, Oliver and Cooper, Nick J and Quinlan, Aaron R and Mychaleckyj, Josyf C and Farber, Emily and Bonnie, Jessica K and Szpak, Michal and Schofield, Ellen and others},
  journal={Nature genetics},
  volume={47},
  number={4},
  pages={381--386},
  year={2015},
  publisher={Nature Publishing Group}
}

@article{barban2016genome,
  title={Genome-wide analysis identifies 12 loci influencing human reproductive behavior},
  author={Barban, Nicola and Jansen, Rick and De Vlaming, Ronald and Vaez, Ahmad and Mandemakers, Jornt J and Tropf, Felix C and Shen, Xia and Wilson, James F and Chasman, Daniel I and Nolte, Ilja M and others},
  journal={Nature genetics},
  volume={48},
  number={12},
  pages={1462--1472},
  year={2016},
  publisher={Nature Publishing Group US New York}
}

@article{malik2018multiancestry,
  title={Multiancestry genome-wide association study of 520,000 subjects identifies 32 loci associated with stroke and stroke subtypes},
  author={Malik, Rainer and Chauhan, Ganesh and Traylor, Matthew and Sargurupremraj, Muralidharan and Okada, Yukinori and Mishra, Aniket and Rutten-Jacobs, Loes and Giese, Anne-Katrin and Van Der Laan, Sander W and Gretarsdottir, Solveig and others},
  journal={Nature genetics},
  volume={50},
  number={4},
  pages={524--537},
  year={2018},
  publisher={Nature Publishing Group US New York}
}

@article{otowa2016meta,
  title={Meta-analysis of genome-wide association studies of anxiety disorders},
  author={Otowa, Takeshi and Hek, Karin and Lee, Minyoung and Byrne, Enda M and Mirza, Saira S and Nivard, Michel G and Bigdeli, Timothy and Aggen, Steven H and Adkins, Daniel and Wolen, Aaron and others},
  journal={Molecular psychiatry},
  volume={21},
  number={10},
  pages={1391--1399},
  year={2016},
  publisher={Nature Publishing Group}
}

@article{locke2015genetic,
  title={Genetic studies of body mass index yield new insights for obesity biology},
  author={Locke, Adam E and Kahali, Bratati and Berndt, Sonja I and Justice, Anne E and Pers, Tune H and Day, Felix R and Powell, Corey and Vedantam, Sailaja and Buchkovich, Martin L and Yang, Jian and others},
  journal={Nature},
  volume={518},
  number={7538},
  pages={197--206},
  year={2015},
  publisher={Nature Publishing Group UK London}
}

@article{liu2019association,
  title={Association studies of up to 1.2 million individuals yield new insights into the genetic etiology of tobacco and alcohol use},
  author={Liu, Mengzhen and Jiang, Yu and Wedow, Robbee and Li, Yue and Brazel, David M and Chen, Fang and Datta, Gargi and Davila-Velderrain, Jose and McGuire, Daniel and Tian, Chao and others},
  journal={Nature genetics},
  volume={51},
  number={2},
  pages={237--244},
  year={2019},
  publisher={Nature Publishing Group US New York}
}

@article{day2018elucidating,
  title={Elucidating the genetic basis of social interaction and isolation},
  author={Day, Felix R and Ong, Ken K and Perry, John RB},
  journal={Nature communications},
  volume={9},
  number={1},
  pages={2457},
  year={2018},
  publisher={Nature Publishing Group UK London}
}

@article{wray2018genome,
  title={Genome-wide association analyses identify 44 risk variants and refine the genetic architecture of major depression},
  author={Wray, Naomi R and Ripke, Stephan and Mattheisen, Manuel and Trzaskowski, Maciej and Byrne, Enda M and Abdellaoui, Abdel and Adams, Mark J and Agerbo, Esben and Air, Tracy M and Andlauer, Till MF and others},
  journal={Nature genetics},
  volume={50},
  number={5},
  pages={668--681},
  year={2018},
  publisher={Nature Publishing Group US New York}
}

@article{jansen2019genome,
  title={Genome-wide analysis of insomnia in 1,331,010 individuals identifies new risk loci and functional pathways},
  author={Jansen, Philip R and Watanabe, Kyoko and Stringer, Sven and Skene, Nathan and Bryois, Julien and Hammerschlag, Anke R and de Leeuw, Christiaan A and Benjamins, Jeroen S and Mu{\~n}oz-Manchado, Ana B and Nagel, Mats and others},
  journal={Nature genetics},
  volume={51},
  number={3},
  pages={394--403},
  year={2019},
  publisher={Nature Publishing Group US New York}
}

@article{nelson2017association,
  title={Association analyses based on false discovery rate implicate new loci for coronary artery disease},
  author={Nelson, Christopher P and Goel, Anuj and Butterworth, Adam S and Kanoni, Stavroula and Webb, Tom R and Marouli, Eirini and Zeng, Lingyao and Ntalla, Ioanna and Lai, Florence Y and Hopewell, Jemma C and others},
  journal={Nature genetics},
  volume={49},
  number={9},
  pages={1385--1391},
  year={2017},
  publisher={Nature Publishing Group US New York}
}

@article{zengini2018genome,
  title={Genome-wide analyses using UK Biobank data provide insights into the genetic architecture of osteoarthritis},
  author={Zengini, Eleni and Hatzikotoulas, Konstantinos and Tachmazidou, Ioanna and Steinberg, Julia and Hartwig, Fernando P and Southam, Lorraine and Hackinger, Sophie and Boer, Cindy G and Styrkarsdottir, Unnur and Gilly, Arthur and others},
  journal={Nature genetics},
  volume={50},
  number={4},
  pages={549--558},
  year={2018},
  publisher={Nature Publishing Group US New York}
}

@article{michailidou2017association,
  title={Association analysis identifies 65 new breast cancer risk loci},
  author={Michailidou, Kyriaki and Lindstr{\"o}m, Sara and Dennis, Joe and Beesley, Jonathan and Hui, Shirley and Kar, Siddhartha and Lema{\c{c}}on, Audrey and Soucy, Penny and Glubb, Dylan and Rostamianfar, Asha and others},
  journal={Nature},
  volume={551},
  number={7678},
  pages={92--94},
  year={2017},
  publisher={Nature Publishing Group UK London}
}

@article{demenais2018multiancestry,
  title={Multiancestry association study identifies new asthma risk loci that colocalize with immune-cell enhancer marks},
  author={Demenais, Florence and Margaritte-Jeannin, Patricia and Barnes, Kathleen C and Cookson, William OC and Altm{\"u}ller, Janine and Ang, Wei and Barr, R Graham and Beaty, Terri H and Becker, Allan B and Beilby, John and others},
  journal={Nature genetics},
  volume={50},
  number={1},
  pages={42--53},
  year={2018},
  publisher={Nature Publishing Group US New York}
}

@article{schumacher2018association,
  title={Association analyses of more than 140,000 men identify 63 new prostate cancer susceptibility loci},
  author={Schumacher, Fredrick R and Al Olama, Ali Amin and Berndt, Sonja I and Benlloch, Sara and Ahmed, Mahbubl and Saunders, Edward J and Dadaev, Tokhir and Leongamornlert, Daniel and Anokian, Ezequiel and Cieza-Borrella, Clara and others},
  journal={Nature genetics},
  volume={50},
  number={7},
  pages={928--936},
  year={2018},
  publisher={Nature Publishing Group US New York}
}

@article{wood2014defining,
  title={Defining the role of common variation in the genomic and biological architecture of adult human height},
  author={Wood, Andrew R and Esko, Tonu and Yang, Jian and Vedantam, Sailaja and Pers, Tune H and Gustafsson, Stefan and Chu, Audrey Y and Estrada, Karol and Luan, Jian'an and Kutalik, Zolt{\'a}n and others},
  journal={Nature genetics},
  volume={46},
  number={11},
  pages={1173--1186},
  year={2014},
  publisher={Nature Publishing Group US New York}
}

@article{hoffmann2018large,
  title={A large electronic-health-record-based genome-wide study of serum lipids},
  author={Hoffmann, Thomas J and Theusch, Elizabeth and Haldar, Tanushree and Ranatunga, Dilrini K and Jorgenson, Eric and Medina, Marisa W and Kvale, Mark N and Kwok, Pui-Yan and Schaefer, Catherine and Krauss, Ronald M and others},
  journal={Nature genetics},
  volume={50},
  number={3},
  pages={401--413},
  year={2018},
  publisher={Nature Publishing Group US New York}
}

@article{deelen2019meta,
  title={A meta-analysis of genome-wide association studies identifies multiple longevity genes},
  author={Deelen, Joris and Evans, Daniel S and Arking, Dan E and Tesi, Niccol{\`o} and Nygaard, Marianne and Liu, Xiaomin and Wojczynski, Mary K and Biggs, Mary L and van Der Spek, Ashley and Atzmon, Gil and others},
  journal={Nature communications},
  volume={10},
  number={1},
  pages={3669},
  year={2019},
  publisher={Nature Publishing Group UK London}
}

@article{klimentidis2018genome,
  title={Genome-wide association study of habitual physical activity in over 377,000 UK Biobank participants identifies multiple variants including CADM2 and APOE},
  author={Klimentidis, Yann C and Raichlen, David A and Bea, Jennifer and Garcia, David O and Wineinger, Nathan E and Mandarino, Lawrence J and Alexander, Gene E and Chen, Zhao and Going, Scott B},
  journal={International journal of obesity},
  volume={42},
  number={6},
  pages={1161--1176},
  year={2018},
  publisher={Nature Publishing Group UK London}
}

@article{karlsson2019genome,
  title={Genome-wide association analyses of risk tolerance and risky behaviors in over 1 million individuals identify hundreds of loci and shared genetic influences},
  author={Karlsson Linn{\'e}r, Richard and Biroli, Pietro and Kong, Edward and Meddens, S Fleur W and Wedow, Robbee and Fontana, Mark Alan and Lebreton, Ma{\"e}l and Tino, Stephen P and Abdellaoui, Abdel and Hammerschlag, Anke R and others},
  journal={Nature genetics},
  volume={51},
  number={2},
  pages={245--257},
  year={2019},
  publisher={Nature Publishing Group US New York}
}

@article{jansen2019alzh,
  title={Genome-wide meta-analysis identifies new loci and functional pathways influencing Alzheimer’s disease risk},
  author={Jansen, Iris E and Savage, Jeanne E and Watanabe, Kyoko and Bryois, Julien and Williams, Dylan M and Steinberg, Stacy and Sealock, Julia and Karlsson, Ida K and H{\"a}gg, Sara and Athanasiu, Lavinia and others},
  journal={Nature genetics},
  volume={51},
  number={3},
  pages={404--413},
  year={2019},
  publisher={Nature Publishing Group US New York}
}
\bibliographystyle{naturemag}
\renewcommand{\bibnumfmt}[1]{#1}


\clearpage
\pagebreak
\begin{appendix}
\part*{Appendix}

\appendix
\renewcommand{\thefigure}{\thesection.\arabic{figure}}
\renewcommand{\thetable}{\thesection.\arabic{table}}
\setcounter{figure}{0}
\setcounter{table}{0}
\makeatletter
\@addtoreset{figure}{section}
\@addtoreset{table}{section}
\makeatother

\section{Extended Results}\label{sec:coef45}
\begin{figure}[!htbp]
    \centering
    \includegraphics[width=0.9\textwidth]{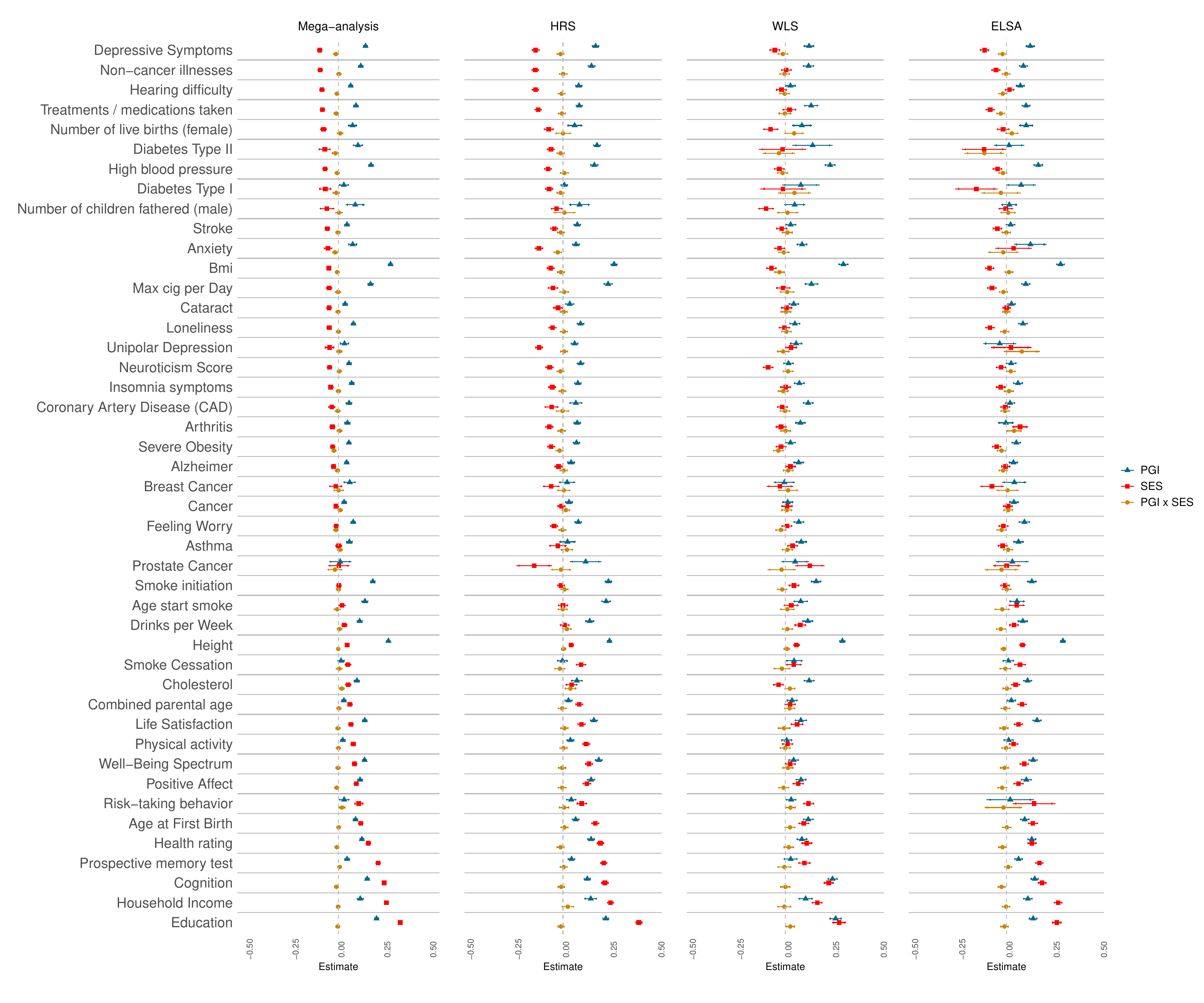}
    \caption{Associations between PGIs, parental SES, and PGI $\times$ SES across phenotypes}
    \justifying
    \vspace{0.2cm}
    \noindent \footnotesize{Plot of the coefficients estimated from the OLS regression model specified in \autoref{eq:OLS_gxe}, showing the associations, along with their 90\% and 95\% confidence intervals, between polygenic indices (PGIs), parental socioeconomic status (SES), and their interaction (PGI $\times$ SES) across 45 phenotypes. The estimates are obtained from a mega-analysis that pools all three datasets---the HRS, the WLS, and ELSA---and from separate analyses for each individual database. Control variables include an indicator for each dataset, 10 ancestry-specific principal components, age and age squared, sex, region of birth, and interactions between age and sex.}
    \label{fig:ols_gxses_pgs1_all1}
\end{figure}

Table \ref{tab:coef45} reports the estimates and standard errors for the PGI, parental SES, and the PGI $\times$ SES interaction across the 45 phenotypes, obtained from the mega-analysis pooling HRS, WLS, and ELSA.
{\footnotesize
\setlength{\tabcolsep}{6pt}
\renewcommand{\arraystretch}{0.95}
\begin{longtable}{lrrr}
\caption{Associations between PGIs, parental SES, and PGI $\times$ SES across phenotypes}
\label{tab:coef45} \\
\toprule
Phenotype & PGI ($\beta^s$) & SES ($\gamma^s$) & PGI $\times$ SES ($\rho^s$) \\ 
\midrule
\endfirsthead
\multicolumn{4}{c}{\tablename\ \thetable\ -- \textit{Continued from previous page}} \\
\toprule
Phenotype & PGI ($\beta^s$) & SES ($\gamma^s$) & PGI $\times$ SES ($\rho^s$) \\ 
\midrule
\endhead
\midrule
\multicolumn{4}{r}{\textit{Continued on next page}} \\
\endfoot
\bottomrule
\endlastfoot
Depressive Symptoms & 0.148*** (0.007) & -0.103*** (0.007) & -0.014* (0.007) \\ 
Non-cancer illnesses & 0.122*** (0.007) & -0.099*** (0.007) & 0.002 (0.007) \\ 
Hearing difficulty & 0.067*** (0.007) & -0.090*** (0.007) & -0.008 (0.007) \\ 
Treatments / medications taken & 0.095*** (0.007) & -0.088*** (0.007) & -0.011 (0.007) \\ 
Number of live births (female) & 0.077*** (0.013) & -0.082*** (0.009) & 0.010 (0.009) \\ 
Diabetes Type II & 0.106*** (0.014) & -0.074*** (0.017) & -0.016 (0.010) \\ 
High blood pressure & 0.177*** (0.007) & -0.072*** (0.007) & -0.007 (0.007) \\ 
Diabetes Type I & 0.029* (0.014) & -0.072*** (0.018) & -0.012 (0.010) \\ 
Number of children fathered (male) & 0.091*** (0.026) & -0.063** (0.021) & 0.003 (0.011) \\ 
Stroke & 0.047*** (0.007) & -0.061*** (0.007) & -0.003 (0.007) \\ 
Anxiety & 0.077*** (0.013) & -0.057*** (0.012) & -0.018* (0.008) \\ 
BMI & 0.284*** (0.007) & -0.052*** (0.007) & -0.007 (0.007) \\ 
Max cig per Day & 0.175*** (0.008) & -0.051*** (0.009) & -0.002 (0.008) \\ 
Cataract & 0.037*** (0.007) & -0.051*** (0.007) & -0.002 (0.007) \\ 
Loneliness & 0.082*** (0.007) & -0.050*** (0.007) & 0.000 (0.007) \\ 
Unipolar Depression & 0.033* (0.013) & -0.049*** (0.013) & 0.004 (0.009) \\ 
Neuroticism Score & 0.057*** (0.007) & -0.048*** (0.007) & 0.004 (0.007) \\ 
Insomnia symptoms & 0.072*** (0.007) & -0.042*** (0.007) & -0.000 (0.007) \\ 
Coronary Artery Disease (CAD) & 0.058*** (0.010) & -0.037*** (0.010) & -0.003 (0.008) \\ 
Arthritis & 0.049*** (0.008) & -0.033*** (0.008) & 0.005 (0.008) \\ 
Severe Obesity & 0.057*** (0.007) & -0.031*** (0.007) & -0.024*** (0.007) \\ 
Alzheimer & 0.045*** (0.007) & -0.027*** (0.007) & -0.005 (0.007) \\ 
Breast Cancer & 0.061*** (0.017) & -0.015 (0.019) & 0.001 (0.015) \\ 
Cancer & 0.031*** (0.007) & -0.013 (0.007) & 0.010 (0.007) \\ 
Feeling Worry & 0.080*** (0.007) & -0.013 (0.007) & -0.013 (0.007) \\ 
Asthma & 0.059*** (0.009) & 0.001 (0.009) & 0.009 (0.008) \\ 
Prostate Cancer & 0.010 (0.031) & 0.003 (0.030) & -0.019 (0.021) \\ 
Smoke initiation & 0.187*** (0.007) & 0.003 (0.007) & 0.001 (0.007) \\ 
Age start smoke & 0.144*** (0.010) & 0.019 (0.010) & -0.007 (0.010) \\ 
Drinks per Week & 0.115*** (0.008) & 0.032*** (0.008) & 0.005 (0.008) \\ 
Height & 0.272*** (0.005) & 0.048*** (0.005) & -0.001 (0.005) \\ 
Smoke Cessation & 0.014 (0.010) & 0.051*** (0.009) & 0.004 (0.009) \\ 
Cholesterol & 0.100*** (0.009) & 0.053*** (0.009) & 0.018* (0.008) \\ 
Combined parental age & 0.030*** (0.007) & 0.062*** (0.007) & 0.002 (0.007) \\ 
Life Satisfaction & 0.143*** (0.007) & 0.068*** (0.007) & -0.003 (0.007) \\ 
Physical activity & 0.023** (0.007) & 0.081*** (0.007) & 0.000 (0.007) \\ 
Well-Being Spectrum & 0.142*** (0.007) & 0.087*** (0.007) & -0.002 (0.007) \\ 
Positive Affect & 0.117*** (0.008) & 0.098*** (0.007) & -0.009 (0.007) \\ 
Risk-taking behavior & 0.031* (0.015) & 0.111*** (0.013) & 0.019* (0.010) \\ 
Age at First Birth & 0.093*** (0.007) & 0.122*** (0.007) & 0.002 (0.007) \\ 
Health rating & 0.128*** (0.007) & 0.163*** (0.007) & -0.008 (0.007) \\ 
Prospective memory test & 0.047*** (0.008) & 0.216*** (0.007) & 0.007 (0.007) \\ 
Cognition & 0.157*** (0.007) & 0.249*** (0.007) & -0.011 (0.007) \\ 
Household Income & 0.119*** (0.010) & 0.261*** (0.007) & -0.002 (0.007) \\ 
Education & 0.207*** (0.006) & 0.336*** (0.007) & -0.003 (0.006) \\ 
\end{longtable}
}

\justifying
\footnotesize Note: Estimated coefficients from the OLS regression model specified in \autoref{eq:OLS_gxe}, showing the associations between polygenic indices (PGIs), parental socioeconomic status (SES), and their interaction (PGI $\times$ SES) across 45 phenotypes, for the mega-analysis pooling HRS, WLS, and ELSA. Standard errors are reported in parentheses, and phenotypes are sorted in ascending order of the estimated parental SES coefficient, matching the ordering of Figure~\ref{fig:ols_gxses_pgs1_all}. Control variables include an indicator for each dataset, 10 ancestry-specific principal components, age and age squared, sex, region of birth, and interactions between age and sex. Significance: $^{*}\,p<0.05$; $^{**}\,p<0.01$; $^{***}\,p<0.001$.

\pagebreak

\section{Phenotype selection} \label{sec:GWASs}
\subsection{GWAS catalog and criteria for selection}
To select the most relevant and powerful GWASs for this study we used the following methodology. First, we consulted the GWAS Catalog (\url{https://www.ebi.ac.uk/gwas/downloads}; version of 30 May 2019) \cite{GWASCatalog_2013}, which contained 776 GWASs. Second, we divided all GWASs into 11 major trait families based on the literature \cite{watanabe2019global}. Third, we used the following criteria to filter out GWASs not suited for our study:
\begin{itemize}
    \item Low genome-wide significant SNP count (Association count $<$ 6).
    \item More relevant trait available.
    \item Low prevalence.
    \item Not in HRS.
    \item Non-European sample.
    \item Larger GWAS (for this trait) exists.
    \item Other study/measure used.
\end{itemize}
\vspace{5mm}

We made the filter steps quantifiable when possible. However, in determining the relevance of a specific trait to our study we had to rely on subjective assessment. We labeled a trait as ``More relevant GWAS available'' when, for instance, instead of ``coronary artery disease, type I diabetes mellitus'' we used the main ``coronary artery disease'' and main ``type I diabetes'' phenotype GWAS.

Given that we had three datasets to test the PGI in, the trait also had to be present in those datasets. Therefore, for this initial filtering we filtered out traits when they were not in the HRS, by labeling them ``Not in HRS''. This filter step was based on a Google search of the terms ``Health and Retirement Study'' in combination with the trait.

\subsection{Selection methodology in the GWAS Catalog}
After the first filtering step leaving only potentially relevant GWASs, we conducted a rigorous selection process. All remaining GWASs were subjected to a comparison within their respective trait families. Generally, we included the GWASs of a unique trait with the largest sample size within a given trait domain. For instance, we included the largest GWASs within the domain `Metabolic', which were, among others, diabetes type 2 and BMI. The power of a GWAS was based on the sample size and the GWAS association count. Furthermore, we included GWASs with lower GWAS association counts if these GWASs provided interesting phenotypic information beyond the information provided by the most powerful GWASs within a trait family. For example, we included the phenotype `obesity' because we assessed that this phenotype provides information on top of what `BMI' provides.

Furthermore, since we also consulted the GWAS Atlas (discussed in the next section), we also assessed whether the selected GWAS from the GWAS Catalog was a better choice than those available from the GWAS Atlas.

\subsection{Selection methodology in the GWAS Atlas}
We consulted the GWAS Atlas (Watanabe et al. (2019)\cite{watanabe2019global}; \url{https://atlas.ctglab.nl/}, version of 17 April 2019). Regarding filter and selection steps, we used similar principles as described in the previous sections. However, since the GWAS Atlas provides more detailed information regarding the GWASs, we were able to apply more rigorous filtering steps. The filtering criteria for the GWAS Atlas were:
\begin{itemize}
    \item Number of SNPs $>$ 450,000.
    \item SNP-based heritability Z-score $>$ 2.
    \item Contains EUR population.
    \item Low genome-wide significant SNP count (Association count $<$ 6).
    \item Selected nonbinary traits with $>$50,000 European individuals with non-missing phenotypes; selected binary traits for which available cases and controls were both $>$10,000 and the total sample size was $>$50,000.
\end{itemize}

Similar to the selection process in the GWAS Catalog, we selected the largest and most relevant GWASs of the trait domains as indicated by the GWAS Atlas \cite{watanabe2019global}. Finally, we compared the selected GWASs from the GWAS Atlas with those of the GWAS Catalog. The final choice regarding inclusion was documented and based on the sample size of the GWAS and its relevance, and assessed independently by two researchers.

Note that, in addition to published GWASs, the GWAS Atlas also contained GWASs specifically conducted for the GWAS Atlas project \cite{watanabe2019global}. We prioritized previously published GWASs and only included GWAS-Atlas-specific GWASs if the phenotype was important to our study and there was no equivalent previously published GWAS available in the GWAS Catalog or GWAS Atlas.

\subsection{Processing of selected GWASs}
Subsequently, summary statistics of the GWASs selected from the GWAS Atlas and GWAS Catalog were downloaded. The summary statistics that had no information regarding the number of SNPs and SNP-h2 Z-score were put through a final round of filtering using LDSC \cite{bulik2015ld}, to ensure that only well-powered GWASs were included with a sufficient number of SNPs. Next, we included only those phenotypes that are available in the HRS, ELSA, and WLS.

After downloading and reviewing the summary statistics and reading the papers of the selected GWASs, we made the following minor adaptations:

\begin{itemize}

\item We revised the selection of phenotypes to improve consistency and data availability. Specifically, the social interaction phenotype was replaced with a loneliness phenotype derived from the same study, as both summary statistics and corresponding variables are available.

\item The phenotype subjective well-being was replaced by four related constructs---positive affect, neuroticism, life satisfaction, and depressive symptoms---based on the multivariate GWAS by Baselmans et al. (2019)\cite{baselmans2019multivariate}.

\item We excluded the phenotypes unipolar depression and Major Depressive Disorder 2 (MDD2) from Wray et al. (2018)\cite{wray2018genome}, as more recent and comprehensive GWAS results are available \cite{howard2019genome, baselmans2019multivariate}.

\item For parental longevity, we retained only the combined parental age at death (z-scored). Alternative measures (e.g., father's age at death, mother's age at death, and indicators of extreme longevity) were excluded due to redundancy. The combined measure was preferred because it maximizes sample size (by including individuals with only one deceased parent) and exhibits the highest explanatory power ($R^2$).

\item The phenotype extreme/severe obesity was excluded due to duplication with the obese vs.\ thin phenotype, as both definitions are effectively equivalent. Additionally, a continuous BMI measure is already included.

\item For general risk-taking behavior, we use the more recent GWAS summary statistics from Karlsson Linnér et al. (2019)\cite{karlsson2019genome}.

\item For age at first birth, we use the pooled GWAS summary statistics rather than sex-specific estimates, as this maximizes sample size and maintains consistency with our main analyses, which are not stratified by gender.

\item Regarding subjective well-being and its related constructs (e.g., neuroticism and life satisfaction), we rely on multivariate GWAS results rather than univariate summary statistics.

\item We established the following criteria for selecting variables in the Health and Retirement Study (HRS):
\begin{itemize}
    \item For general intelligence, we prioritize the cognition score (cognitionScore) over the Mental Status Index (mentalStatus) and Word Recall Index (wordRecall), provided that missingness is not excessive.
\end{itemize}

\item When exact phenotype matches are unavailable in HRS, we use proxy variables:
\begin{itemize}
    \item General risk-taking behavior: we use behavioral measures (e.g., gambling games), if available, rather than self-reported risk preferences, due to closer conceptual alignment with GWAS measures.

    \item Sleep-related traits: GWAS measures are based on sleep duration (hours slept), which is not directly available in HRS. Instead, we use survey questions capturing early awakening and perceived restfulness. However, given the availability of an insomnia variable, these proxies are excluded from the final analysis.

    \item Physical activity: GWAS measures of moderate-to-vigorous physical activity (MVPA) are based on weighted weekly minutes of activity. In HRS, we construct a binary indicator (actModVig) equal to 1 if respondents engage in moderate or vigorous physical activity at least once per week, and 0 otherwise.
\end{itemize}

\item As a general rule for selecting GWAS results, we prioritize multivariate methods such as MTAG when available, as they increase statistical power while distinguishing between related phenotypes. However, we do not use Genomic SEM or similar approaches that produce latent-factor GWAS, in order to maintain interpretability of individual phenotypes.

\end{itemize}

In total, our final selection includes 45 GWASs. A complete overview of the included studies is provided in Supplementary Table \ref{tab:gwas_references}.

{\footnotesize
\renewcommand{\arraystretch}{0.95}
\begin{longtable}{p{0.42\textwidth} p{0.48\textwidth}}
\caption{GWAS references for phenotypes used in the analysis}
\label{tab:gwas_references} \\
\toprule
Phenotype & Reference \\
\midrule
\endfirsthead
\multicolumn{2}{c}{\tablename\ \thetable\ -- \textit{Continued from previous page}} \\
\toprule
Phenotype & Reference \\
\midrule
\endhead
\midrule
\multicolumn{2}{r}{\textit{Continued on next page}} \\
\endfoot
\bottomrule
\endlastfoot
Age at First Birth & Barban et al. (2016)\cite{barban2016genome} \\
Age start smoke & Liu et al. (2019)\cite{liu2019association} \\
Alzheimer & Jansen et al. (2019)\cite{jansen2019alzh} \\
Anxiety & Otowa et al. (2016)\cite{otowa2016meta} \\
Arthritis & Zengini et al. (2018)\cite{zengini2018genome} \\
Asthma & Demenais et al. (2018)\cite{demenais2018multiancestry} \\
BMI & Locke et al. (2015)\cite{locke2015genetic} \\
Breast Cancer & Michailidou et al. (2017)\cite{michailidou2017association} \\
Cancer & Neale Lab UKB (2018)\cite{NealeUKB2018} \\
Cataract & Neale Lab UKB (2018)\cite{NealeUKB2018} \\
Cholesterol & Hoffmann et al. (2018)\cite{hoffmann2018large} \\
Cognition & Savage et al. (2018)\cite{savage2018genome} \\
Combined parental age & Deelen et al. (2019)\cite{deelen2019meta} \\
Coronary Artery Disease (CAD) & Nelson et al. (2017)\cite{nelson2017association} \\
Depressive Symptoms & Baselmans et al. (2019)\cite{baselmans2019multivariate} \\
Diabetes Type I & Onengut-Gumuscu et al. (2015)\cite{onengut2015fine} \\
Diabetes Type II & Mahajan et al. (2018)\cite{mahajan2018fine} \\
Drinks per Week & Liu et al. (2019)\cite{liu2019association} \\
Education & Lee et al. (2018)\cite{lee2018gene} \\
Feeling Worry & Otowa et al. (2016)\cite{otowa2016meta} \\
Health rating & Neale Lab UKB (2018)\cite{NealeUKB2018} \\
Hearing difficulty & Neale Lab UKB (2018)\cite{NealeUKB2018} \\
Height & Wood et al. (2014)\cite{wood2014defining} \\
High blood pressure & Neale Lab UKB (2018)\cite{NealeUKB2018} \\
Household Income & Neale Lab UKB (2018)\cite{NealeUKB2018} \\
Insomnia symptoms & Jansen et al. (2019)\cite{jansen2019genome} \\
Life Satisfaction & Baselmans et al. (2019)\cite{baselmans2019multivariate} \\
Loneliness & Day et al. (2018)\cite{day2018elucidating} \\
Max cig per day & Liu et al. (2019)\cite{liu2019association} \\
Neuroticism Score & Baselmans et al. (2019)\cite{baselmans2019multivariate} \\
Non-cancer illnesses & Neale Lab UKB (2018)\cite{NealeUKB2018} \\
Number of children fathered (male) & Barban et al. (2016)\cite{barban2016genome} \\
Number of live births (female) & Neale Lab UKB (2018)\cite{NealeUKB2018} \\
Physical activity & Klimentidis et al. (2018)\cite{klimentidis2018genome} \\
Positive Affect & Baselmans et al. (2019)\cite{baselmans2019multivariate} \\
Prostate Cancer & Schumacher et al. (2018)\cite{schumacher2018association} \\
Prospective memory test & Neale Lab UKB (2018)\cite{NealeUKB2018} \\
Risk-taking behavior & Karlsson Linnér et al. (2019)\cite{karlsson2019genome} \\
Severe Obesity & Locke et al. (2015)\cite{locke2015genetic} \\
Smoke cessation & Liu et al. (2019)\cite{liu2019association} \\
Smoke initiation & Liu et al. (2019)\cite{liu2019association} \\
Stroke & Malik et al. (2018)\cite{malik2018multiancestry} \\
Treatments / medications taken & Neale Lab UKB (2018)\cite{NealeUKB2018} \\
Unipolar Depression & Wray et al. (2018)\cite{wray2018genome} \\
Well-Being Spectrum & Baselmans et al. (2019)\cite{baselmans2019multivariate} \\
\bottomrule
\end{longtable}
}

\section{Data Cleaning: HRS} \label{sec:HRSclean}
We constructed most phenotypes using RAND HRS Longitudinal files. RAND HRS Longitudinal Files collect, clean, and rename variables from Core and Exit Interviews of the HRS. Other variables were extracted directly from the yearly study datasets. In this case, the variables used to create the phenotypes are not necessarily present in all rounds.

Some general data cleaning was required on HRS variables. Afterwards, we merged a panel with all waves through a unique identifier.

\subsection{Height (height)}
The height variable is present in all rounds and is part of the RAND files (\textbf{rheight}).
\subsection{Body Mass Index (bmi)}
We extract the BMI variable directly from the RAND files. Values over 70 were set to missing, since they are considered to be data entry errors.
\subsection{Obese vs.\ thin (obesitySevere)}
We created the extreme obesity variable based on the definition from the UKB GWAS. Severe obesity is a dummy variable equal to 1 if BMI is greater than or equal to 40 $kg/m^2$, and 0 otherwise.
\subsection{General intelligence latent factor (cognitionScore)}
The total cognition score is calculated by HRS as the sum of the Total Recall Index and the Mental Status Index.

The total recall index (RwTR20, R1TR40, RwATR20, RwHTR40) is available in all waves and summarizes the immediate and delayed word recall tasks. In Waves 1 and 2H, the recall wordlist contained 20 words, while in all other waves it contained 10. Thus the scores range from 0 to 40 in Waves 1 and 2H and from 0 to 20 in other waves.

The mental status index (RwMSTOT, R2AMSTOT) sums scores from counting, naming, and vocabulary tasks and is available for Wave 2A and from Wave 3 forward. This reflects the absence of some of these tests in Waves 1 and 2H.

A total cognition score (RwCOGTOT, R2ACGTOT) sums the total recall and mental status indices. Because the mental status index is missing for Waves 1 and 2H, the total cognition index is also missing for these waves. The total cognition score ranges from 0 to 35.

Further detail on these variables can be found in the RAND HRS Longitudinal File Codebook.
\subsection{Life satisfaction (lifeSatisfaction)}
Life satisfaction variables are not included in RAND files and were extracted directly from the leave-behind section from 2004 onwards. Variables extracted are \textbf{LB003A--LB003E}.

We constructed the life satisfaction variable based on \cite{Smith2017} by averaging the scores across all 5 items from the life satisfaction section. We set the final score to missing if there are three or more items with missing values, as indicated in the document.

We recoded 2006 variables since for that year the response scale was 6-point and changed to a 7-point scale thereafter.
\subsection{Positive affect (positiveAffect)}
Positive affect variables are not included in RAND files and were extracted directly from the leave-behind section from 2006 onwards. Variables extracted are \textbf{LB027A--LB027Y}.

We constructed the positive affect variable based on \cite{Smith2017} by reverse-coding items c, d, f, g, h, k, p, q, t, u, v, x, and y, and averaging the scores across all 13 items. We set the final score to missing if there are more than six items with missing values.
\subsection{Neuroticism (neuroticismScore)}
Calculation of the neuroticism score is based on \cite{Lachman1997} and \cite{Smith2017}.

Neuroticism variables are not included in RAND files and were extracted directly from the leave-behind section from 2006 onwards. Variables extracted are \textbf{LB033A--LB033Z}.

We calculate neuroticism by reverse-coding questions d, h, l, q, and averaging the scores across the 4 items. We set the final score to missing if more than half of the items have missing values within each sub-dimension.
\subsection{Depressive symptoms (depressScore)}
For depressive symptoms we use RAND variable \textbf{RwCESD}.

The CESD score (\textbf{RwCESD}) is the sum of five ``negative'' indicators minus two ``positive'' indicators. The negative indicators measure whether the Respondent experienced the following sentiments all or most of the time: depression, everything is an effort, sleep is restless, felt alone, felt sad, and could not get going. The positive indicators measure whether the Respondent felt happy and enjoyed life all or most of the time. The corresponding RAND variable is \textbf{rcesd}.
\subsection{Worry measurement, neuroticism factor (worryFeeling)}
We constructed feeling worry from the question ``Please indicate how well each of the following describes you: Worrying'', with answers ``1. A LOT, 2. SOME, 3. A LITTLE, 4. NOT AT ALL''.
\subsection{General risk-taking behavior (risk)}
We calculated risk aversion based on \cite{Kimball2008}, supplemental material section. Hypothetical gambles on lifetime income are available for waves 1992 and 1994, and 1998 to 2006. The questions change through rounds, therefore we created 3 measures that can be merged.

The HRS has fielded two versions, the first or ``original'' version and the revised or ``status-quo bias free'' version. The original questions begin:
\begin{itemize}
    \item ``Suppose that you are the only income earner in the family, and you have a good job guaranteed to give you your current (family) income every year for life. You are given the opportunity to take a new and equally good job, with a 50-50 chance it will double your (family) income and a 50-50 chance that it will cut your (family) income by a third. Would you take the new job?''
\end{itemize}
The revised questions (bias-free) begin:
\begin{itemize}
    \item ``Suppose that you are the only income earner in the family. Your doctor recommends that you move because of allergies, and you have to choose between two possible jobs. The first would guarantee your current total family income for life. The second is possibly better paying, but the income is also less certain. There is a 50-50 chance the second job would double your total lifetime income and a 50-50 chance that it would cut it by a third. Which job would you take---the first job or the second job?''
\end{itemize}

\textbf{rc4\_o1}: Wave 1992 had only 2 gambles on the question, therefore only 4 categories are possible. Categories are defined from high risk aversion (rejects all gambles) to low risk aversion (accepts all gambles). This wave is susceptible to ``status quo'' bias.

\textbf{rc6\_f}: Wave 1994 had 4 gambles on the question, therefore 6 categories are possible. Categories are defined from high risk aversion (rejects all gambles) to low risk aversion (accepts all gambles). This wave is susceptible to ``status quo'' bias.

\textbf{rc6\_f}: From waves 1998 to 2006, the question had 4 gambles, therefore 6 categories are possible. Categories are defined from high risk aversion (rejects all gambles) to low risk aversion (accepts all gambles). This wave is free of ``status quo'' bias.
\subsection{Loneliness (loneliness)}
We measured loneliness based on \cite{Smith2017}. HRS incorporates a 3-item and an 11-item scale of loneliness derived from the 20-item Revised UCLA Loneliness Scale.

We construct the loneliness measure by reverse-coding questions a, b, c, and e. Then we averaged the scores across all 11 items. We set the final score to missing if more than five items have missing values.

To create the original 3-item loneliness index, we reverse-code items a, b, c, and create an average of these three scores. We set the final score to missing if more than 1 item is missing.
\subsection{Insomnia symptoms (insomniaFrequent)}
Insomnia questions are present from 2002 to 2016. To create the variable for insomnia symptoms, we used the questions ``How often have trouble with waking up during the night?'' and ``How often have trouble falling asleep?''. Possible answers were ``1. Most of the time, 2. Sometimes, 3. Rarely or never''.

Variable \textbf{insomnia} is coded 1 if the person responded ``Most of the time'' or ``Sometimes'' to one of the two questions.

Variable \textbf{insomniaFrequent} is coded 1 if the person responded ``Most of the time'' to one of the two questions. Our selected variable is \textbf{insomniaFrequent}.

HRS variables 2002--2016:\\
\textbf{(H, J, K, L, M, N, O, P)083}: How often do you have trouble falling asleep---would you say most of the time, sometimes, or rarely or never?\\
\textbf{(H, J, K, L, M, N, O, P)084}: How often do you have trouble with waking up during the night---would you say most of the time, sometimes, or rarely or never?
\subsection{Unipolar depression (depress)}
We constructed unipolar depression from the question ``Felt depressed'' from the CES-D questionnaire. RAND variables are \textbf{rdeprex} for wave 1 and \textbf{rdepres} for all other waves.
\subsection{Anxiety (anxietyScore)}
Five items were selected from the widely used Beck Anxiety Inventory (BAI). The Beck Inventory has been shown to distinguish symptoms of anxiety from depression and to be valid for use in older populations. This scale was not included after 2012.

\begin{table}[htbp]
  \centering
  \caption{5 items (Q41a--Q41e)}
    \begin{tabular}{ll}
    \toprule
    Q41a & I had fear of the worst happening. \\
    Q41b & I was nervous. \\
    Q41c & I felt my hands trembling. \\
    Q41d & I had a fear of dying. \\
    Q41e & I felt faint. \\
    \bottomrule
    \end{tabular}
  \label{tab:Anxiety_BAI}
\end{table}

Scaling: Responses to the 5 items are averaged to form an index of anxiety ranging from 1--4 (1: never feeling x to 4: most of the time feeling). Set the final score to missing if more than two of the items have missing values.
\subsection{Alcohol consumption, drinks per week (drinkWeek)}
RAND variables used to calculate the number of drinks per week are: drinks per day (\textbf{rdrinkn} and \textbf{rdrinkr}) and number of days per week drinking (\textbf{rdrinkd}).
\subsection{Smoking initiation (ageSmoke)}
Age at smoking initiation is constructed from the question ``How old were you when you first started smoking cigarettes?'', taking the earliest reported age across all available waves.
\subsection{Smoking cessation (cesSmoke\_GSCAN)}
Following the phenotype definition used in the GSCAN consortium\cite{GSCAN2019gwas}, smoking cessation is coded as 1 for former smokers and 0 for current smokers, restricting the sample to ever-smokers. Constructed from RAND variable \textbf{rsmoken}, which indicates whether the Respondent currently smokes.
\subsection{Number of cigarettes smoked per day (cigsAll\_Current)}
This variable was constructed from individual file data using the questions ``NUM CIGARETTES SMOKED PER DAY'' and ``NUM CIGS PER DAY---WHEN SMOKED MOST''.

\begin{table}[htbp]
  \centering
  \caption{Variables from individual files HRS}
    \begin{tabular}{rll}
    \toprule
    Year & Num cig smoked per day & Num cig smoked per day, when smoked most \\
    \midrule
    1992 & V503  & V505 \\
    1994 & W453  &  \\
    1996 & E943  &  \\
    1998 & F1268 & F1275 \\
    2000 & F1401 & G1408 \\
    2002 & HC118 & HC123 \\
    2004 & JC118 & JC123 \\
    2006 & KC118 & KC123 \\
    2008 & LC118 & LC123 \\
    2010 & MC118 & MC123 \\
    2012 & NC118 & NC123 \\
    2014 & OC118 & OC123 \\
    \bottomrule
    \end{tabular}
  \label{tab:cpd}
\end{table}
\subsection{Physical activity measurement (actModVig)}
RAND data on physical activity start from 2006 to 2016. We have created the variable \textbf{actModVig} that takes value 1 for those who do moderate or vigorous physical activity once a week or more often, and 0 otherwise.

Vigorous physical activity: ``We would like to know the type and amount of physical activity involved in your daily life. How often do you take part in sports or activities that are vigorous, such as running or jogging, swimming, cycling, aerobics or gym workout, tennis, or digging with a spade or shovel: more than once a week, once a week, one to three times a month, or hardly ever or never?''

Moderate physical activity: ``And how often do you take part in sports or activities that are moderately energetic, such as gardening, cleaning the car, walking at a moderate pace, dancing, floor or stretching exercises: more than once a week, once a week, one to three times a month, or hardly ever or never?''
\subsection{Parental longevity (ageParentsZ)}
Variables for parental longevity were created from \textbf{rmomage}, \textbf{rdadage}, \textbf{rmomliv}, \textbf{rdadliv}. We took the age of alive parents or at parent's death.

The 1\% oldest cut-off for HRS was 99 for women and 96 for men.
\subsection{Osteoarthritis (arthritis)}
Osteoarthritis is a variable from RAND files. RAND files contain the questions ``R reports arthritis/rheumatism this wave'' and ``Whether or not a doctor has ever told the Respondent s/he had arthritis or rheumatism'' (\textbf{rarthr}, \textbf{rarthre}).
\subsection{Type II diabetes (diabetes)}
Type II diabetes is a variable from RAND files. RAND files contain the questions ``R reports diabetes this wave'' and ``Respondent has ever reported having diabetes'' (\textbf{rdiab}, \textbf{rdiabe}).
\subsection{Educational attainment (educYears)}
Educational attainment in HRS is constructed from the question ``R Years of Education'' (\textbf{raedyrs}).
\subsection{Age at first birth, AFB (ageFirstBirth)}
Constructed using \textbf{HwAGEOKID} variables from RAND Family data files, renamed as \textbf{ageOldch}, the age of the Respondent's oldest child. This variable is derived from the best-guess child's age (\textbf{KwAGEBG}) in the Respondent-kid file. Some of the ages are over 80 years old; these ages are based on the reported birth year.

Some cleaning was required for this variable since there were cases where the age of the oldest child was greater than the age of the respondent. These cases were coded as missing values.

\begin{table}[htbp]
  \centering
  \caption{Inconsistent values in oldest child's age}
    \begin{tabular}{lrrr}
    \toprule
     & Freq. & Percent & Cum. \\
    \midrule
    Respondent is older & 485{,}244 & 99.84 & 99.84 \\
    Child is older & 780 & 0.16 & 100.00 \\
    \midrule
    Total & 486{,}024 & 100.00 & \\
    \bottomrule
    \end{tabular}
  \label{tab:Missingsageolderchild}
\end{table}

We found cases where the difference between the respondent and the oldest child is less than 5 years or greater than 60. These inconsistencies depend on the reported age of the oldest child based on the respondent's best guess. In addition, we coded as missing those cases where the age at first birth is over 60 years old for women.

\begin{table}[htbp]
  \centering
  \caption{Age at first birth from RAND family data}
    \begin{tabular}{lrrrrr}
    \toprule
    Variable & Obs & Mean & Std. Dev. & Min & Max \\
    \midrule
    ageFirstBirth & 207{,}782 & 23.23 & 6.22 & 0 & 79 \\
    \bottomrule
    \end{tabular}
  \label{tab:sumAFB}
\end{table}

To account for these inconsistent values in the age at first birth (\textbf{ageFirstBirth}), the variable \textbf{FlagAFB} flags values under 5 for the full sample and over 60 for women, where childbearing is extremely unlikely. We replace these flagged values with missing.
\subsection{Total cholesterol measurement (totChol)}
The total cholesterol variable was extracted from the 2016 Venous Blood Study. The variable is measured in mg/dl.
\subsection{Breast carcinoma (cancerBreast)}
This variable is included only in years 1992 and 1994. The variable is coded as 1 if breast cancer was diagnosed since the last round (yes/no).
\subsection{Prostate carcinoma (cancerProstate)}
This variable is included only in years 1992 and 1994. The variable is coded as 1 if prostate cancer was diagnosed since the last round (yes/no).

\section{Data Cleaning: WLS} \label{sec:WLSclean}
WLS private data contain 30 duplicated observations. Of these, 28 observations are pairs of identical twins. We delete one observation per twin pair. We merge genetic data by unique ID and keep a sample of White respondents.

\subsection{Height (height)}
Measures of height in WLS are self-reported for rounds 4 to 6. Round 6 also includes a variable for height measured in person, which we use. Height is measured in inches, and we convert it to centimeters to match the GWAS definition.
\subsection{General intelligence latent factor (cognitionScore)}
There are multiple IQ measures, but \textbf{gwiiq\_bm} is the WLS recommended one if only one measure is to be used for WLS Graduates.

Cognitive ability (IQ) was measured in the freshman and junior year of high school (1956) using the Henmon--Nelson test of mental ability. The Henmon--Nelson test is a group-administered, 30-minute, multiple-choice assessment that consists of 90 verbal or quantitative items. It was administered in all Wisconsin high schools at various grade levels from the 1930s through the 1960s as part of a cooperative effort of high schools and colleges to identify youth who might succeed in college but did not originally plan to attend college.

Instead of using the best measure recommended by WLS, some studies also use the raw score. Raw Henmon--Nelson test scores were converted to IQ scores by standardizing to a mean of 100 based on centile rank. IQ scores ranged from 61 to 145; for an example see \cite{Stephan2018}.

There are other variables that measure some form of cognition. In waves 1992, 2004, and 2011, WLS included a subset of the fourteen items from the Wechsler Adult Intelligence Scale (WAIS). In wave 2004, cognition was measured with a fluency section and a word recall section.

\begin{table}[htbp]
  \centering
  \caption{Cognition variables: Wechsler Adult Intelligence Scale (WAIS)}
    \begin{tabular}{ll}
    \toprule
    ri001re & R4 total cognition score \\
    z\_gi101re & R5 6-item score for cognition \\
    z\_gi106re & R5 9-item score for cognition \\
    z\_hi101re & R6 6-item score summarizing R's performance on selected WAIS questions \\
    \bottomrule
    \end{tabular}
  \label{tab:cogWAIS}
\end{table}
\subsection{Life satisfaction (lifeSatisfaction)}
There is no direct measure of life satisfaction in WLS. The closest measure is psychological well-being, captured through eudaimonia. Questions are asked about six different aspects of life (autonomy, environmental mastery, personal growth, positive relations with others, purpose in life, and self-acceptance), with the answers combined to form a composite eudaimonic measure. Further explanation on how to construct this variable can be found in \cite{WLSScale}. The continuous variable was constructed from 12 items in 1992--1993 and 18 items in 2003--2005.

\begin{table}[htbp]
  \centering
  \caption{Psychological well-being: 12-item}
    \begin{tabular}{ll}
    \toprule
    z\_rn016rec & I am influenced by people with strong opinions \\
    z\_rn017rec & I am in charge of how I live \\
    z\_rn018rec & It is hard to maintain close relations \\
    z\_rn019rec & I do not wander aimlessly through life \\
    z\_rn020rec & I am pleased with how things turned out \\
    z\_rn021rec & Demands of everyday life often get me down \\
    z\_rn022rec & Life is a continuous process of learning, changing, growing \\
    z\_rn023rec & Have not experienced many warm, trusting relationships \\
    z\_rn024rec & Live life one day at a time, do not think about the future \\
    z\_rn025rec & I judge myself by what I think is important \\
    z\_rn026rec & Gave up trying to make big improvements/changes in life \\
    z\_rn027rec & I like most aspects of my personality \\
    \bottomrule
    \end{tabular}
  \label{tab:Wellbeing}
\end{table}

\begin{table}[htbp]
  \centering
  \caption{Psychological well-being: 18-item}
    \begin{tabular}{ll}
    \toprule
    gn116re & I tend to be influenced by people with strong opinions. \\
    gn125re & I judge myself by what I think is important, not by what others think. \\
    gn128re & I have confidence in my own opinions even if they are contrary to general consensus. \\
    gn117re & In general, I feel I am in charge of the situation in which I live. \\
    gn121re & The demands of everyday life often get me down. \\
    gn129re & I am quite good at managing the many responsibilities of my daily life. \\
    gn122re & For me, life has been a continuous process of learning, changing, and growing. \\
    gn126re & I gave up trying to make big improvements or changes in my life a long time ago. \\
    gn130re & It is important to have new experiences that challenge how I think about myself/world. \\
    gn118re & Maintaining close relationships has been difficult and frustrating for me. \\
    gn123re & I have not experienced many warm and trusting relationships with others. \\
    gn131re & People would describe me as a giving person, willing to share my time with others. \\
    gn119re & Some people wander aimlessly through life, but I am not one of them. \\
    gn124re & I live life one day at a time and do not really think about the future. \\
    gn132re & I sometimes feel as if I have done all there is to do in life. \\
    gn120re & When I look at the story of my life, I am pleased with how things have turned out. \\
    gn127re & I like most aspects of my personality. \\
    gn133re & In many ways, I feel disappointed about my achievements in life. \\
    \bottomrule
    \end{tabular}
  \label{tab:wellbeing}
\end{table}
\subsection{Positive affect (positiveAffect)}
There is no direct measure of positive affect in the WLS questionnaire. However, WLS measures emotion level based on the Health Utilities Index (HUI). This variable is constructed from 3 items:

\begin{table}[htbp]
  \centering
  \caption{Emotion score questions}
    \begin{tabular}{ll}
    \toprule
    z\_gx330re & During the past 4 weeks, have you been feeling happy or unhappy? \\
    z\_gx331re & Felt happy and interested in life, or somewhat happy? \\
    z\_gx332re & Felt somewhat unhappy, very unhappy, or so unhappy life not worthwhile? \\
    \bottomrule
    \end{tabular}
  \label{tab:EmotionLevel}
\end{table}
\subsection{Neuroticism (neuroticismScore)}
We create neuroticism as described in the scales documentation of WLS \cite{WLSScale}.

\begin{table}[htbp]
  \centering
  \caption{Neuroticism variables}
    \begin{tabular}{ll}
    \toprule
    z\_mh025rec & R4 measure of neuroticism \\
    z\_ih025rec & R5 summary score for neuroticism \\
    z\_jh025rec & R6 summary score for neuroticism \\
    \bottomrule
    \end{tabular}
  \label{tab:neuroticisml}
\end{table}
\subsection{Depressive symptoms (depressScore)}
The methodology to calculate the depression score with a 20-item scale is described in \cite{Savard1957}. For this scale we used variables \textbf{z\_mu001rec}, \textbf{z\_iu001rec}, and \textbf{z\_ju001rec}.

\begin{table}[htbp]
  \centering
  \caption{20-item depression variables}
    \begin{tabular}{ll}
    \toprule
    z\_mu001rec & Summary score for psychological distress/depression---modified CES-D \\
    z\_iu001rec & Summary score for psychological distress/depression---modified CES-D \\
    z\_ju001rec & Summary score for psychological distress/depression---modified CES-D \\
    \bottomrule
    \end{tabular}
  \label{tab:depressScore_WLS1}
\end{table}

To replicate the 8-item score used in other surveys, we used the following questions:

\begin{table}[htbp]
  \centering
  \caption{8-item depression score}
    \begin{tabular}{ll}
    \toprule
    z\_ju013rer & R6 On how many days during the past week did you feel depressed? \\
    z\_ju012rer & R6 On how many days during the past week did you feel sad? \\
    z\_ju008rer & R6 On how many days during the past week did you feel lonely? \\
    z\_ju017rer & R6 How many days in the past week felt that everything you did was an effort? \\
    z\_ju020rer & R6 On how many days during the past week did you sleep restlessly? \\
    z\_ju022rer & R6 On how many days during the past week did you feel you could not `get going'? \\
    z\_ju006rer & R6 On how many days during the past week did you feel happy? \\
    z\_ju009rer & R6 On how many days during the past week did you enjoy life? \\
    \bottomrule
    \end{tabular}
  \label{tab:depressionScore8_WLS2}
\end{table}

The preferred measure is the 20-item score.
\subsection{Worry measurement, neuroticism factor (worryFeeling)}
Feeling worry was constructed from the question ``On how many days in the past week did you worry over possible misfortune?''.

\begin{table}[htbp]
  \centering
  \caption{Feeling worry questions}
    \begin{tabular}{ll}
    \toprule
    nu041rer & R4 days past week worry over possible misfortune \\
    z\_iu041rer & R5 On how many days during the past week did you worry over possible misfortune? \\
    z\_ju041rer & R6 On how many days in the past week did you worry over possible misfortune? \\
    \bottomrule
    \end{tabular}
  \label{tab:WorryFeeling}
\end{table}
\subsection{General risk-taking behavior (risk)}
To measure risk we used the question ``Importance of having a low risk of losing your job vs.\ high pay'' (\textbf{z\_mg004rer}, \textbf{z\_ig004rer}, \textbf{z\_jfin05re}, \textbf{z\_jfin10re}).
\subsection{Loneliness (loneliness)}
Loneliness is measured with the question ``In a week how lonely are you (1--4)'' (\textbf{z\_ik016rer}). This is only asked in round 5.
\subsection{Insomnia symptoms (insomniaFrequent)}
We constructed \textbf{insomniaFrequent} from variables \textbf{mx020rer} and \textbf{z\_ix020rer}, which ask: ``How often have you had trouble sleeping in the past 6 months?''. \textbf{insomniaFrequent} is coded 1 if the person has sleeping problems more than once a week.
\subsection{Unipolar depression (depress)}
We constructed unipolar depression based on the question ``Did R ever feel sad, blue, or depressed for two weeks or longer?'' from variables \textbf{z\_gu002re}, \textbf{z\_ru002re}, and \textbf{z\_au002re}.
\subsection{Anxiety (anxietyScore)}
As explained in \cite{WLSScale}, WLS uses the Spielberger State-Trait Anxiety Inventory (STAI) to assess anxiety. It can be used in clinical settings to diagnose anxiety and to distinguish it from depressive syndromes. It is also often used in research as an indicator of caregiver distress.

The index is constructed from survey questions:

\begin{table}[htbp]
  \centering
  \caption{Anxiety questions: On how many days during the past week did you...}
    \begin{tabular}{ll}
    \toprule
    IU035RER & ``feel calm?'' \\
    IU037RER & ``feel tense?'' \\
    IU039RER & ``feel at ease?'' \\
    IU041RER & ``worry over possible misfortune?'' \\
    IU043RER & ``feel nervous?'' \\
    IU045RER & ``feel jittery?'' \\
    IU047RER & ``feel relaxed?'' \\
    \bottomrule
    \end{tabular}
  \label{tab:Anxiety_WLS}
\end{table}

The index was constructed by summing the valid values across the seven items if at least five items received a valid response. The variable was coded as $-2$ if fewer than five items received a valid response. Variables IU035RER, IU039RER, and IU047RER were reverse-coded in the creation of this variable.
\subsection{Alcohol consumption, drinks per week (drinkWeek)}
Drinks per week was constructed from the question ``Total number of drinks in the last month'' for rounds 4, 5, and 6 (variables \textbf{z\_ru028re}, \textbf{z\_gu028re}, \textbf{z\_hu028re}). We divided the drinks in the last month by 30 and multiplied by 7. The final value is the maximum drinks per week across the three rounds.
\subsection{Smoking initiation (ageSmoke)}
Age at smoking initiation is constructed from the question ``How old were you when you started smoking cigarettes regularly?'', asked in rounds 5 and 6. We take the minimum reported age across rounds.
\subsection{Smoking cessation (cesSmoke\_GSCAN)}
Following the phenotype definition used in the GSCAN consortium\cite{GSCAN2019gwas}, smoking cessation is coded as 1 for former smokers and 0 for current smokers, restricting the sample to ever-smokers. Constructed using current smoking status across rounds 4, 5, and 6 from variables \textbf{z\_mx013rer}, \textbf{z\_ix013rec}, and \textbf{z\_jx013rec}.
\subsection{Number of cigarettes smoked per day (cigsAll\_Current)}
Maximum cigarettes per day was constructed from the question ``On average, how many packs of cigarettes do you smoke a day?'' in rounds 4, 5, and 6 (variables \textbf{nx044red}, \textbf{z\_ixt08rer}, \textbf{z\_jxt08rer}). We use the maximum value across rounds.
\subsection{Physical activity measurement (actModVig)}
This survey asks only about light physical activity. The question is ``Past year, how many hours per month did you do light physical activities with others?'' for rounds 5 and 6 (variables \textbf{z\_jz168rer}, \textbf{z\_iz168rer}). We converted hours per month to minutes per week by multiplying by 60 and dividing by 4.3452 (the average number of weeks per month). We use the maximum value across rounds.
\subsection{Parental longevity (ageParentsZ)}
We calculated parental longevity from parent age (if alive) and parent's age at death. We standardized parental age.
\subsection{Osteoarthritis (arthritis)}
We calculated this variable using the question ``Has a medical professional ever said you have arthritis or rheumatism?'', present in rounds 4, 5, and 6 (variables \textbf{z\_mx087rer}, \textbf{z\_gx360re}, \textbf{z\_jx211rer}). We code yes if the respondent has ever been told they have arthritis.
\subsection{Type II diabetes (diabetes)}
We calculated this variable using the question ``Has a medical professional ever said you have diabetes?'', present in rounds 4, 5, and 6 (variables \textbf{z\_mx095rer}, \textbf{z\_gx342re}, \textbf{z\_hx342re}). We code yes if the respondent has ever been told they have diabetes.
\subsection{Educational attainment (educYears)}
Years of education in WLS is constructed from \textbf{edhi}, indicating the highest level of education attained, converted into years using standard mappings (e.g., high school = 12, bachelor's = 16, master's = 18, doctoral = 21). For respondents with multiple reports across waves, we use the highest reported value.
\subsection{Age at first birth, AFB (ageFirstBirth)}
Age at first birth is asked only in round 3. If the age at first birth is missing, we replace it with a constructed age at first birth, calculated by subtracting the age of the oldest biological child from the age of the respondent.
\subsection{Total cholesterol measurement (totChol)}
There is no direct measure of cholesterol in WLS. We used the question ``Has a medical professional ever said that you have high cholesterol?'' to indicate high total cholesterol.
\subsection{Breast carcinoma (cancerBreast)}
Breast cancer is measured with the question ``Has had breast cancer'' (\textbf{z\_gx350ere}). This question is present only in round 5.
\subsection{Prostate carcinoma (cancerProstate)}
Prostate cancer is measured with the question ``Has had prostate cancer'' (\textbf{z\_gx351ere}). This question is present only in round 5.

\section{Data Cleaning: ELSA}\label{sec:ELSclean}

\subsection{Height (height)}
Constructed from \textbf{htval}.
\subsection{Obese vs.\ thin (obesitySevere)}
We created the extreme obesity variable based on the definition from the UKB GWAS. Severe obesity is a dummy variable equal to 1 if BMI is greater than or equal to 40 $kg/m^2$, and 0 otherwise.
\subsection{Body Mass Index (bmi)}
We extract the BMI measure directly from the survey (\textbf{bmival}). We replace extreme values (over 70 $kg/m^2$) with missing values.

\begin{table}[htbp]
  \centering
  \caption{Values over 70 BMI}
    \begin{tabular}{lrrr}
    \toprule
    over70 & Freq. & Percent & Cum. \\
    \midrule
    0 & 23{,}179 & 95.14 & 95.14 \\
    1 & 1{,}184 & 4.86 & 100.00 \\
    \midrule
    Total & 24{,}363 & 100.00 & \\
    \bottomrule
    \end{tabular}
  \label{tab:over70}
\end{table}
\subsection{General intelligence latent factor (cognitionScore)}
Based on \cite{Gale2014}, we construct general intelligence as the sum of 3 standardized cognition tests. These tests are verbal fluency, immediate and delayed verbal memory, and attention. Verbal (semantic) fluency was assessed by asking participants to name as many animals as they could think of in 1 minute. Immediate and delayed verbal memory was assessed by presenting a list of 10 nouns aurally on a computer, one every 2 seconds. Participants were asked to recall as many words as possible immediately and again after a short delay during which they carried out the other cognitive tests. Attention and mental speed were assessed using a letter cancellation task. Participants were given a clipboard to which was attached a page of 780 random letters of the alphabet set out in a grid of 26 rows and 30 columns, and were asked to cross out as many target letters (P and W) as possible in 1 minute.

The cognition score is the average of standardized verbal fluency, immediate and delayed verbal memory, and attention.

\begin{table}[htbp]
  \centering
  \caption{Questions on cognition}
    \begin{tabular}{ll}
    \toprule
    cflisen & Number of words recalled immediately \\
    cflisd & Number of words recalled after delay \\
    cfani & Number of animals mentioned (fluency) \\
    nrowcl & Letter cancellation task: total number of letters searched ($30 \times (\text{nrow}-1) + \text{nclm}$) \\
    \bottomrule
    \end{tabular}
  \label{tab:cog}
\end{table}
\subsection{Life satisfaction (lifeSatisfaction)}
Life satisfaction questions are included from wave 2 onwards.

\begin{table}[htbp]
  \centering
  \caption{Satisfaction with life}
    \begin{tabular}{ll}
    \toprule
    sclifea & In most ways his/her life is close to his/her ideal \\
    sclifeb & The conditions of his/her life are excellent \\
    sclifec & Is satisfied with his/her life \\
    sclifed & So far, he/she has got the important things he/she wants in life \\
    sclifee & If could live his/her life again, would change almost nothing \\
    \bottomrule
    \end{tabular}
  \label{tab:lifeSatisfaction}
\end{table}

We constructed the life satisfaction variable based on \cite{Diener1985} by averaging the scores across all 5 items from the life satisfaction section. We set the final score to missing if there are three or more items with missing values, as indicated in the document.
\subsection{Positive affect (positiveAffect)}
Positive affect variables are included in the main questionnaire only for wave 5. Variables from the positive affect section are \textbf{scfede, scfeen, scfeac, scfepr, scfeint, scfeha, scfeat, scfeco, scfeins, scfeho, scfeal, scfeca,} and \textbf{scfeex}.

We constructed the positive affect variable based on \cite{Steptoe2013} by reverse-coding all questions and averaging the scores across all 13 items. We set the final score to missing if there are more than six items with missing values.
\subsection{Neuroticism (neuroticismScore)}
Calculation of the neuroticism score is based on \cite{Lachman1997}. Personality traits questions are:

\begin{table}[ht]
  \centering
  \caption{Personality traits questions}
   \resizebox{0.5\textwidth}{!}{
    \begin{tabular}{ll}
    \toprule
    \multicolumn{2}{c}{How well does the following describe the respondent?} \\
    \midrule
    scdewa & warm \\
    scdewo & worrying \\
    scdere & responsible \\
    scdeli & lively \\
    scdeca & caring \\
    scdene & nervous \\
    scdecr & creative \\
    scdeha & hardworking \\
    scdeim & imaginative \\
    scdesof & soft-hearted \\
    scdecal & calm \\
    scdein & intelligent \\
    scdecu & curious \\
    scdeac & active \\
    scdecar & careless \\
    scdebr & broad-minded \\
    scdesy & sympathetic \\
    scdeta & talkative \\
    scdeso & sophisticated \\
    scdead & adventurous \\
    scdeth & thorough \\
    \bottomrule
    \end{tabular}
   }
  \label{tab:big5}
\end{table}

For the neuroticism variable we used \textbf{scdemo, scdewo, scdene, scdecal}.
\subsection{Depressive symptoms (depressScore)}
To construct the depression score we use \cite{VandeVelde2009} and \cite{Karim2015}. We construct the variable based on the methodology presented for HRS, since the 8-item scale is the same as the one used in HRS by \cite{Steffick2000}.

The CESD score (\textbf{depressScore}) is the sum of five ``negative'' indicators minus two ``positive'' indicators. The negative indicators measure whether the Respondent experienced the following sentiments all or most of the time: depression, everything is an effort, sleep is restless, felt alone, felt sad, and could not get going. The positive indicators measure whether the Respondent felt happy and enjoyed life all or most of the time. If more than half of the questions are missing, the score is set to missing.

\begin{table}[htbp]
  \centering
  \caption{Variables used for depression}
    \begin{tabular}{ll}
    \toprule
    psceda & Whether felt depressed much of the time during past week \\
    pscedb & Whether felt everything they did during past week was an effort \\
    pscedc & Whether felt their sleep was restless during past week \\
    pscedd & Whether was happy much of the time during past week \\
    pscede & Whether felt lonely much of the time during past week \\
    pscedf & Whether enjoyed life much of the time during past week \\
    pscedg & Whether felt sad much of the time during past week \\
    pscedh & Whether could not get going much of the time during past week \\
    \bottomrule
    \end{tabular}
  \label{tab:depressScore_HRS}
\end{table}
\subsection{Worry measurement, neuroticism factor (worryFeeling)}
We constructed feeling worry based on the question \textbf{scdewo} from the personality trait questionnaire.
\subsection{General risk-taking behavior (risk)}
The risk module is only included in wave 5. Question \textbf{ririsk} asks ``Whether prepared to take risks on a scale of 0 to 10''.
\subsection{Loneliness (loneliness)}
ELSA measures loneliness with a four-item scale \cite{Hughes2004}, which is based on the widely-used 20-item Revised UCLA Loneliness Scale. A fifth item was added in round 3.

We created both a 4-item and a 5-item loneliness scale. We construct the loneliness measure by reverse-coding questions a, b, c, and e, then averaging the scores across all 4 or 5 items. We set the final score to missing if more than 2 items are missing.

In addition, we added a variable called \textbf{lonely} that asks: ``How often respondent feels lonely''.
\subsection{Insomnia symptoms (insomniaFrequent)}
We created insomnia symptoms from the questions ``Sleep: how often respondent has difficulty falling asleep'' (\textbf{heslpa}) and ``Sleep: frequency wake up several times at night'' (\textbf{heslpb}). Frequent insomnia is a dummy variable equal to 1 if the person had difficulty falling asleep and woke up several times a night more than once a week, and 0 otherwise.
\subsection{Unipolar depression (depress)}
Unipolar depression is constructed from the question ``Psychiatric problem has: depression'' (\textbf{hepsyde}).
\subsection{Anxiety (anxietyScore)}
We constructed anxiety from the question \textbf{hepsyan} for round 3 onward and from questions \textbf{hepsy1, hepsy2, hepsy3, hepsy4, hepsy5, hepsy6} for rounds 1 and 2.
\subsection{Parental longevity (ageParentsZ)}
Variables for parental longevity were created from \textbf{dianm}, \textbf{dianf}, \textbf{dinma}, \textbf{dinfa}, \textbf{mthagd}, \textbf{fthagd}, \textbf{dimad}, \textbf{difad}. We took the age of alive parents or at parent's death. Ages under 10 years old were set to missing.
\subsection{Osteoarthritis (arthritis)}
Osteoarthritis is created from variables \textbf{heart1}, \textbf{heart2}, \textbf{heart3}, \textbf{heartoa}.

\begin{table}[htbp]
  \centering
  \caption{Arthritis questions}
    \begin{tabular}{ll}
    \toprule
    heart1 & Which types of arthritis do you have? 1st \\
    heart2 & Which types of arthritis do you have? 2nd \\
    heart3 & Which types of arthritis do you have? 3rd \\
    heartoa & Whether has osteoarthritis \\
    \bottomrule
    \end{tabular}
  \label{tab:arthritis}
\end{table}
\subsection{Type II diabetes (diabetes)}
Diabetes is calculated from variables \textbf{dheacd}, \textbf{heacd}, ``Ever been told had diabetes (from feed or doctor)''.
\subsection{Educational attainment (educYears)}
Years of education is calculated from the variable ``age finished continuous full-time education, merged from current and previous waves'' (\textbf{edend}).
\subsection{Age at first birth, AFB (ageFirstBirth)}
We calculated age at first birth by subtracting the age of the oldest natural son or daughter from the reported age of the respondent at the moment of the survey.

\textbf{Age of the oldest natural child:} We calculated the age of the oldest child by taking the maximum age across reported ages (\textbf{chage1--chage16}) for natural children only (\textbf{chtype1--chtype16}). However, the age of the oldest child appears inconsistent in some cases.

\textbf{Age at first birth:} Age at first birth is calculated as the difference in age between the respondent and the oldest child in each round. We found some inconsistencies in the data, including negative ages. To account for these inconsistent values, the variable \textbf{FlagAFB} flags ages under 5 for the full sample and over 60 for women, where childbearing is extremely unlikely. We replace these flagged values with missing.

\begin{table}[htbp]
  \centering
  \caption{Age at first birth (ageFirstBirth)}
    \begin{tabular}{lrrrrr}
    \toprule
    Variable & Obs & Mean & Std. Dev. & Min & Max \\
    \midrule
    ageFirstBirth & 58{,}893 & 26.64 & 5.55 & 5 & 66 \\
    \bottomrule
    \end{tabular}
  \label{tab:AFB}
\end{table}
\subsection{Number of children ever born, NEB (childrenEverBorn)}
Number of children ever born is calculated by counting all natural children of the respondent (\textbf{chtype1--chtype16}) in each round and taking the maximum value across rounds.

\begin{table}[htbp]
  \centering
  \caption{Children ever born}
    \begin{tabular}{lrrrrr}
    \toprule
    Variable & Obs & Mean & Std. Dev. & Min & Max \\
    \midrule
    childrenEverBorn & 7{,}569 & 2.07 & 1.40 & 0 & 13 \\
    \bottomrule
    \end{tabular}
  \label{tab:childrenEverBorn}
\end{table}
\subsection{Total cholesterol measurement (totChol)}
Only available for years 2004, 2008, and 2012.
\subsection{Prostate carcinoma (cancerProstate)}
``Ever had prostate cancer'' is available from 2006 onward.
\subsection{Smoking variables}
All the smoking variables used in the analysis are self-reported:
\begin{enumerate}
    \item Max cigarettes per day (\textit{maxCPD}): Maximum weighted average of cigarettes per day across all waves. Observations below the 1st percentile are set to the 1st percentile and observations above the 99th percentile are set to the 99th percentile.
    \item Ever smoker (\textit{everSmoker}): If the respondent smoked cigarettes at one point in life.
    \item Former smoker (\textit{formerSmoker}): The last non-missing answer the respondent gave was used and categorized either into former smoker or current/never smoker.
    \item Age started smoking (\textit{ageStartSmoke}): Year the respondent started smoking daily.
\end{enumerate}

\section{Quality Control}\label{sec:QC}
We performed extensive quality control procedures on each of our genetic datasets, closely following the steps and thresholds recommended by Marees et al. (2018)\cite{marees2018tutorial}.

\subsection{SNP filtering}
We begin by removing variants with low imputation-quality metrics. When available, we remove all variants with an INFO score (generated by the imputation software IMPUTE2 \cite{akaneya2010ephrin}) below 0.7, using the threshold recommended in \cite{lin2010new}. We then remove all variants which are missing for more than 20\% of samples. After doing the equivalent for samples (described later in the section), we proceed with a more stringent threshold, removing all variants which are missing for more than 2\% of samples. We remove all variants with a minor allele frequency (MAF) $<0.01$ to avoid the inclusion of variants lacking sufficient power to identify phenotypic correlation. In order to eliminate potential genotyping errors, we exclude variants which deviate from Hardy--Weinberg Equilibrium (HWE) with a $p$-value $< 1 \times 10^{-6}$.

\subsection{Sample filtering}
After enforcing the first variant missingness threshold described earlier, we remove all samples missing more than 20\% of remaining variants. We then remove all samples for which we find a mismatch between the sex assigned by the dataset and the sex indicated solely by the genotypic data. To do this, we calculate X-chromosome heterozygosity/homozygosity rates and remove individuals outlying from the cluster of female and male samples. In executing this procedure for HRS and WLS, we raised the threshold for X-chromosome homozygosity for females because the female cluster extended well beyond the recommended threshold of 0.2. After the variant filter process is complete, we remove all samples missing more than 2\% of remaining variants.

We proceed with a heterozygosity rate filter to ensure sample quality and remove examples of possible inbreeding. Considering only a pruned subset of variants, we remove samples for which the heterozygosity rate is more than 3 standard deviations from the mean. Because this paper assumes that each dataset used is composed of unrelated samples, we must also investigate the relatedness of subjects in each dataset. In the case of HRS\cite{weir2012quality} and WLS\cite{WLSScale}, associated quality control reports provide a list of samples to remove in order to achieve a fully ``unrelated'' sample. In the case of ELSA, we follow a procedure roughly similar to that used in the HRS and WLS reports. We calculate the identity by descent (IBD) of all sample pairs and flag all pairs indicated to be second-degree relatives or closer, removing the individual in each pair with the lowest call rate (i.e.\ the highest rate of variant missingness).

The final step of the QC process is to ensure we include only those samples with genotypic data indicating European ancestry by following a detailed population stratification procedure. This process is necessary in order to strengthen the associations we are able to find in our data due to the wide variance of allele frequencies across subpopulations. For our population stratification procedure, we conduct a multidimensional scaling (MDS) analysis on a pruned set of SNPs. Next, we follow the same procedure with data from the 1000 Genomes (1KG) project\cite{10002015global} on a set of SNPs overlapping with our dataset. With these two results, we can visualize the clusters of genetic similarity for each of our datasets, overlaid with that of the 1KG data, for which we know the ethnicity of the samples. We are thus able to isolate the individuals with European ancestry in each of our datasets by including only the individuals overlapping with the European ancestry cluster in the 1KG data.
\end{appendix}

\end{document}